\documentclass[aps,pra,prabib,showpacs,superscriptaddress]{revtex4}%
\usepackage{makeidx}
\usepackage[english]{babel}
\usepackage[latin1]{inputenc}
\usepackage{latexsym}
\usepackage{amssymb}
\usepackage{amsmath}
\usepackage{graphicx}
\usepackage{graphics}
\usepackage{epsfig}
\usepackage{bbm}
\usepackage[dvips]{color}
\usepackage{amsfonts}%
\begin{document}
\title{Superfluid pairing between fermions with unequal masses}
\author{M.A. Baranov}
\affiliation{\mbox{Van der Waals-Zeeman Institute, University of Amsterdam, Valckenierstraat
65, 1018 XE Amsterdam, The Netherlands}}
\affiliation{\mbox{Institut f\"ur Quantenoptik und Quanteninformation,
Technikerstra{\ss}e 21a, A-6020 Innsbruck, Austria}}
\affiliation{\mbox{Russian Research Center Kurchatov Institute, Kurchatov Square 1, 123182 Moscow, Russia}}
\author{C. Lobo}
\affiliation{\mbox{School of Physics and Astronomy, University of Nottingham, University Park, Nottingham,
NG7 2RD, UK}}
\author{G. V. Shlyapnikov}
\affiliation{\mbox{Van der Waals-Zeeman Institute, University of Amsterdam, Valckenierstraat
65, 1018 XE Amsterdam, The Netherlands}}
\affiliation{\mbox{Laboratoire Physique Th\'eorique et Mod\`eles Statistique,
Universit\'e Paris Sud, CNRS, 91405~Orsay, France}}

\begin{abstract}
We consider a superfluid state in a two-component gas of fermionic atoms with
equal densities and unequal masses in the BCS limit. We develop a perturbation
theory along the lines proposed by Gorkov and Melik-Barkhudarov and find that
for a large difference in the masses of heavy ($M$) and light ($m$) atoms one
has to take into account both the second-order and third-order contributions.
The result for the critical temperature and order parameter is then quite
different from the prediction of the simple BCS approach. Moreover, the small
parameter of the theory turns out to be $(p_{F}|a|/\hbar) \sqrt{M/m}\ll1$,
where $p_{F}$ is the Fermi momentum, and $a$ the scattering length. Thus, for
a large mass ratio $M/m$ the conventional perturbation theory requires
significantly smaller Fermi momenta (densities) or scattering lengths than in
the case of $M\sim m$, where the small parameter is $p_{F}|a|/\hbar\ll1$. We
show that $3$-body scattering resonances appearing at a large mass ratio due
to the presence of $3$-body bound Efimov states do not influence the result,
which in this sense becomes universal.

\end{abstract}

\pacs{03.75.Ss, 74.20.Fg  }
\date{\today}
\maketitle

\section{Introduction}

Superfluid pairing in a two-component gas of fermions is a well-known problem
\cite{LL9} lying in the background of extensive studies in condensed matter
and nuclear physics \cite{SC,He3,Migdal}. Recently, this problem was actively
investigated in cold gases of fermionic atoms (see \cite{Trento} for a
review). Experimental efforts were focused on $^{6}$Li or $^{40}$K atoms in
two different internal (hyperfine) states, where one can use Feshbach
resonances for switching the sign and tuning the absolute value of the
interspecies interaction (scattering length $a$), which at resonance changes
from $-\infty$ to $+\infty$. In this respect, one encounters the problem of
BCS-BEC crossover discussed earlier in the context of superconductivity
\cite{Eag,Leg,Noz,Melo,Rand} and for superfluidity of two-dimensional $^{3}$He
films \cite{M,MYu}. On the negative side of the resonance ($a<0$), one should
have the Bardeen-Cooper-Schrieffer (BCS) superfluid pairing at sufficiently
low temperatures, and on the positive side ($a>0$) one expects Bose-Einstein
condensation of diatomic molecules formed by atoms of different components.
Remarkable achievements of cold-atom physics in the last years include the
observation of superfluid behavior through vortex formation in the strongly
interacting regime ($n\left\vert a\right\vert ^{3}\gtrsim1$, where $n$ is the
gas density)\ \cite{zw1}, and the formation and Bose-Einstein condensation of
long-lived weakly bound diatomic molecules at $a>0$ \cite{bec}. Ongoing
experiments with atomic Fermi gases have reached temperatures in the
nanokelvin regime, where at achieved densities one has $T\sim0.1E_{F}$, with
$E_{F}$ being the Fermi energy. For $a<0$ the experiments are now approaching
superfluidity in the BCS limit where $n|a|^{3}\ll1$.

Currently, a new generation of experiments is being set up. In particular, it
is dealing with mixtures of different fermionic atoms or mixtures of fermions
and bosons. The main goal is to reveal the influence of the mass difference on
superfluid properties and to search for novel types of superfluid pairing. The
first experiments demonstrating a possibility of using Feshbach resonances and
creating collisionally stable mixtures of $^{40}$K with $^{6}$Li and/or with
$^{87}$Rb, and $^{6}$Li with $^{23}$Na have already been performed
\cite{Jin,Ket,Ospelkaus,Ing,kai,Jook}. Recent theoretical literature on
mixtures of different fermionic atoms contains a discussion of the BCS limit
\cite{Liu,Caldas,He}, the limit of molecular BEC \cite{PSSJ}, BCS-BEC
crossover \cite{Iskin1,Iskin2,Parish}, and the strongly interacting regime
\cite{Pao}.

In this paper we consider a two-component mixture of fermionic atoms with
different masses and attractive intercomponent interaction in the BCS limit.
It is assumed that the densities of the two species are equal which means that
there is no mismatch between their Fermi surfaces, leading to the usual BCS
type of superfluid pairing. Other kinds of pairing that can occur and compete
with BCS, especially for unequal densities, will be discussed elsewhere
\cite{FFLO}. Here, we generalize the perturbation treatment of the gap
equation, introduced by Gorkov and Melik-Barkhudarov \cite{GMB} for equal
masses of fermions belonging to different components. This approach takes into
account the interaction between the atoms in a Cooper pair due to the
polarization of the medium and allows one to correctly determine the
dependence of the zero-temperature gap $\Delta_{0}$ and superfluid transition
temperature $T_{c}$ on the masses of heavy ($M$) and light ($m$) fermionic
atoms. As we shall see below, already the second order of the perturbation,
the so-called Gorkov-Melik-Barkhudarov contribution, leads to a very different
dependence of the preexponential factor in the expressions for $\Delta_{0}$
and $T_{c}$ on the mass ratio $M/m$, compared to the prediction of the simple
BCS theory.

For a large mass ratio $M/m\gg1$, we include higher order contributions and
show that the actual small parameter of the theory is $(p_{F}|a|/\hbar
)\sqrt{M/m}\ll1$ ($p_{F}$ is the Fermi momentum), not simply $p_{F}%
|a|/\hbar\ll1$ as in the case of $M\sim m$. We give a physical interpretation
of this fact and calculate effective masses of heavy and light fermions.

Large mass ratios $M/m$ are realized in electron-ion plasmas, where the heavy
ion component is usually considered as non-degenerate \cite{LL5}. The
electron-proton pairing in the hydrogen plasma, assuming quantum degeneracy
for both electrons and protons, was discussed by Moulopoulos and Ashcroft
\cite{MA}. They found that at low temperatures the Coulomb electron-proton
attraction leads to the appearance of a (momentum-dependent) gap which for
sufficiently high densities is comparable with the Coulomb interaction at the
mean interparticle separation. Note that this problem is quite different from
ours where the attractive interaction between heavy and light fermions is short-ranged.

Before proceeding with our analysis we make two important remarks. First of
all, if the masses of heavy and light fermionic atoms are very different from
each other and the mass ratio exceeds a critical value, $M/m>13.6$, then two
heavy and one light fermion can form $3$-body weakly bound states. The
appearance of these states, predicted by Efimov \cite{Efimov}, can be easily
understood in the Born-Oppenheimer picture \cite{Fonseca}. If we fix the two
heavy atoms at a relative distance $R<\left\vert a\right\vert $, a localized
state for the light atom appears due to the presence of the heavy pair, which
in turn mediates an attractive interaction $\sim-\hbar^{2}/mR^{2}$ between the
heavy atoms (see, e.g. \cite{PSSJ} and references therein). For a large mass
ratio, $M/m>13.6$, this mediated attraction overcomes the kinetic energy of
the relative motion of the heavy atoms and one has the well-known phenomenon
of ``fall into center" \cite{LL3}. The energy of this state is bounded from
below only due to short-range repulsion. The corresponding wave function of
the relative motion of heavy atoms acquires a large number of nodes thus
showing the presence of many bound states. This makes the $3$-body problem
non-universal in the sense that aside from the $2$-body scattering length $a$,
the description of this problem requires one more parameter - the so-called
$3$-body parameter coming from short-range physics. Also, the presence of
weakly bound Efimov states introduces a resonant character to the $3$-body
scattering problem. This is especially important for the
Gorkov-Melikh-Barkhudarov contribution as it is actually dealing with
processes involving $3$ particles. We, however, have found that the $3$-body
resonances are rather narrow and their contribution is not important. This
makes the Gorkov-Melikh-Barkhudarov approach universal at any mass ratio $M/m$.

Our second remark is related to analogies between BCS pairing of unequal-mass
particles in cold-atom and high energy physics. We wish to emphasize that
there are strong physical differences between the pairing of particles of
different masses in relativistic and in nonrelativistic systems such as cold
atoms. The problem arises in relativistic systems in the study of hadronic
matter \cite{hadron}. It is thought that, at the high densities achieved in
neutron stars, quarks become deconfined and the different types of them (e.g.
up, down and strange quarks) will tend to form Cooper pairs with each other.
These different types of quarks have different masses. This relativistic limit
has been investigated by Kundu and Rajagopal \cite{Kundu} and is characterized
by the Fermi momentum being larger than the bare mass: $p_{F}\gg mc$.
Linearizing the momentum near the Fermi surface $p=p_{F}+\delta p$, ($\delta
p/p_{F}\ll1$), we can expand the free particle energy to the lowest
nonvanishing order in $\delta p/p_{F}$ and $mc/p_{F}$:
\[
E=\sqrt{p^{2}c^{2}+m^{2}c^{4}}\simeq p_{F}c\left(  1+\frac{\delta p}{p_{F}%
}+\frac{m^{2}c^{2}}{2p_{F}^{2}}\right)  .
\]
We see that, as far as the kinetic energy is concerned, a change in mass will
amount to a shift in the chemical potential of the species which is
proportional to $m^{2}$ and depends inversely on $p_{F}$. Therefore, pairing
between particles with different masses in the relativistic limit is
equivalent to studying the problem of pairing of equal mass particles in the
presence of a difference between the chemical potentials of the two species.
In the nonrelativistic limit ($p_{F}\ll mc$) the situation is different:
\[
E\simeq mc^{2}+\frac{p^{2}}{2m}%
\]
and so the mass change cannot be incorporated into the chemical potential,
requiring a very different analysis which we carry out here.

The paper is organized as follows. In Section \ref{sec:bcs} we present general
equations, and in Sections III and IV we calculate the critical temperature
$T_{c}$. Section V is dedicated to the discussion of the small parameter of
the theory, and in Section VI we discuss the order parameter and excitation
spectrum. In Sec. VII we analyze the three-body resonances at a large mass
ratio $M/m$ and show that they do not change the result of the
Gorkov-Melik-Barkhudarov approach. In Sec. VIII we conclude.

\section{General equations}

\label{sec:bcs} We consider a uniform gas composed of heavy and light
fermionic atoms with masses $M$ and $m$, respectively. Both heavy and light
atoms are in a single hyperfine state, and considering low temperatures we
omit heavy-heavy and light-light interactions. The interaction of heavy atoms
with light ones is assumed to be attractive and characterized by a negative
s-wave scattering length $a<0$. The Hamiltonian of the system has the form:%
\begin{equation}
\label{H}H=\int d\mathbf{r}\left[  \sum_{i=1,2}\widehat{\psi}_{i}%
^{+}(\mathbf{r})\left(  -\frac{\hbar^{2}}{2m_{i}}\nabla^{2}-\mu_{i}\right)
\widehat{\psi}_{i}(\mathbf{r})+g\widehat{\psi}_{1}^{+}(\mathbf{r}%
)\widehat{\psi}_{1}(\mathbf{r})\widehat{\psi}_{2}^{+}(\mathbf{r})\widehat
{\psi}_{2}(\mathbf{r})\right]  ,
\end{equation}
where $\widehat{\psi}_{i}(\mathbf{r})$ and $\widehat{\psi}_{i}^{+}$ are the
field operators of fermionic atoms labeled by indices $i=1$ (heavy) and $i=2$
(light), $\mu_{i}$ is the corresponding chemical potential, and $g=2\pi
\hbar^{2}a/m_{r}$ is the coupling constant, with $m_{r}=Mm/(M+m)$ being the
reduced mass. Since the densities of the two species are equal, $n_{1}%
=n_{2}=n$, they have the same Fermi momentum $p_{F}=\hbar(6\pi^{2}n)^{1/3}$,
and, hence, $\mu_{1}=p_{F}^{2}/2M$ and $\mu_{2}=p_{F}^{2}/2m$. Finally, we
require that the system be in the weakly interacting regime, which requires
the inequality $p_{F}|a|/\hbar\ll1 $.

\begin{figure}[ptb]
\begin{center}
\includegraphics[width=10cm]{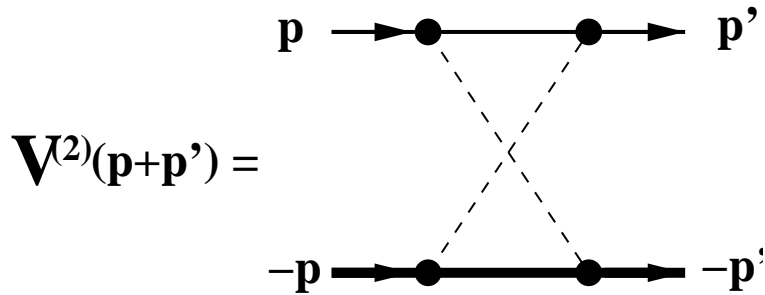}
\end{center}
\caption{The leading many-body contribution (second order, or
Gorkov-Melik-Barkhudarov correction) to the effective interaction between
heavy (thick line) and light (thin line) fermions. The dashed line corresponds
to the coupling constant $g$.}%
\label{Fig1}%
\end{figure}

We now consider the usual BCS scheme where a heavy atom with momentum
$\mathbf{p}$ is paired to a light one having momentum $-\mathbf{p}$. This
leads to the gap equation
\begin{equation}
\Delta(\mathbf{p})=-\int\frac{d\mathbf{p}^{\prime}}{(2\pi\hbar)^{3}%
}V_{\mathrm{eff}}(\mathbf{p},\mathbf{p}^{\prime})\frac{1-f(E_{+}%
(\mathbf{p}^{\prime}))-f(E_{-}(\mathbf{p}^{\prime}))}{E_{+}(\mathbf{p}%
^{\prime})+E_{-}(\mathbf{p}^{\prime})}\Delta(\mathbf{p}^{\prime})
\label{eq:bcsgap}%
\end{equation}
where $f(E)=[\exp(E/T)+1]^{-1}$ is the Fermi-Dirac distribution function, and
we assume that the order parameter is real. The dispersion relations for the
two branches of single-particle excitations are written as
\begin{equation}
E_{\pm}(\mathbf{p})=\pm\left(  \frac{\xi_{1}(\mathbf{p})-\xi_{2}(\mathbf{p}%
)}{2}\right)  +\sqrt{\left(  \frac{\xi_{1}(\mathbf{p})+\xi_{2}(\mathbf{p})}%
{2}\right)  ^{2}+\Delta^{2}}, \label{eq:dispersion}%
\end{equation}
and the quantities $\xi_{1,2}$ are given by $\xi_{1}(\mathbf{p})=(p^{2}%
-p_{F}^{2})/2M$ and $\xi_{2}(\mathbf{p})=(p^{2}-p_{F}^{2})/2m$. The function
$V_{\mathrm{eff}}(\mathbf{p},\mathbf{p}^{\prime})=g+\delta V(\mathbf{p}%
,\mathbf{p}^{\prime})$ is an effective interaction between particles in the
medium, where the quantity $\delta V(\mathbf{p},\mathbf{p}^{\prime})$
originates from many-body effects and is a correction to the bare
interparticle interaction $g$. The leading correction is second order in
\thinspace$g$ and the corresponding diagram is shown in Fig. \ref{Fig1}.

The integral in Eq.~(\ref{eq:bcsgap}) diverges at large momenta due to the
first term in $V_{\mathrm{eff}}$. This divergency can be eliminated by
expressing the bare interaction $g$ in terms of the scattering length $a$
\cite{GMB,LL9,AGD}. If we confine ourselves to the second order in
perturbation theory with respect to $g$, then the renormalized gap equation reads%

\begin{align}
\Delta(\mathbf{p})  &  =-\frac{2\pi\hbar^{2}a}{m_{r}}\int\frac{d\mathbf{p}%
^{\prime}}{(2\pi\hbar)^{3}}\left[  \frac{1-f(E_{+}(\mathbf{p}^{\prime
}))-f(E_{-}(\mathbf{p}^{\prime}))}{E_{+}(\mathbf{p}^{\prime})+E_{-}%
(\mathbf{p}^{\prime})}-\frac{2m_{r}}{p^{\prime2}}\right]  \Delta
(\mathbf{p}^{\prime})\nonumber\\
&  -\int\frac{d\mathbf{p}^{\prime}}{(2\pi\hbar)^{3}}\delta V(\mathbf{p}%
,\mathbf{p}^{\prime})\frac{1-f(E_{+}(\mathbf{p}^{\prime}))-f(E_{-}%
(\mathbf{p}^{\prime}))}{E_{+}(\mathbf{p}^{\prime})+E_{-}(\mathbf{p}^{\prime}%
)}\Delta(\mathbf{p}^{\prime}). \label{eq:renorm_gap}%
\end{align}
The convergence of the integral over $p^{\prime}$ in the first term of the
right-hand side of Eq.~(\ref{eq:renorm_gap}) is now obvious, while the
convergence of the second term is due to the decay of $\delta V$ at large
momenta (see Eq.~(\ref{eq:V2}) below). The gap equation accounting for higher
orders in $g$ will be derived and discussed in Section IV.

In the limit of $\Delta\rightarrow0$ one can reduce Eq.~(\ref{eq:renorm_gap})
to a linearized gap equation:%

\begin{align}
\Delta(\mathbf{p})  &  =-\frac{2\pi\hbar^{2}a}{m_{r}}\int\frac{d\mathbf{p}%
^{\prime}}{(2\pi\hbar)^{3}}\left[  \frac{\tanh[\xi_{1}(\mathbf{p}^{\prime
})/2T]+\tanh[\xi_{2}(\mathbf{p}^{\prime})/2T]}{2[\xi_{1}(\mathbf{p}^{\prime
})+\xi_{2}(\mathbf{p}^{\prime})]}-\frac{2m_{r}}{p^{\prime2}}\right]
\Delta(\mathbf{p}^{\prime})\nonumber\\
&  -\int\frac{d\mathbf{p}^{\prime}}{(2\pi\hbar)^{3}}\delta V(\mathbf{p}%
,\mathbf{p}^{\prime})\frac{\tanh[\xi_{1}(\mathbf{p}^{\prime})/2T]+\tanh
[\xi_{2}(\mathbf{p}^{\prime})/2T]}{2[\xi_{1}(\mathbf{p}^{\prime})+\xi
_{2}(\mathbf{p}^{\prime})]}\Delta(\mathbf{p}^{\prime})
\label{eq:linear_ren_gap}%
\end{align}
The critical temperature $T_{c}$ is determined from
Eq.~(\ref{eq:linear_ren_gap}) as the highest temperature at which this
equation has a non-trivial solution for $\Delta$.

\section{Critical temperature. BCS and GM approaches}

The first line of Eq.~(\ref{eq:linear_ren_gap}) corresponds to the linearized
gap equation in the traditional BCS approach:%

\begin{equation}
\Delta(\mathbf{p})=-\frac{2\pi\hbar^{2}a}{m_{r}}\int\frac{d\mathbf{p}^{\prime
}}{(2\pi\hbar)^{3}}\left[  \frac{\tanh[\xi_{1}(\mathbf{p}^{\prime}%
)/2T]+\tanh[\xi_{2}(\mathbf{p}^{\prime})/2T]}{2[\xi_{1}(\mathbf{p}^{\prime
})+\xi_{2}(\mathbf{p}^{\prime})]}-\frac{2m_{r}}{p^{\prime2}}\right]
\Delta(\mathbf{p}^{\prime}). \label{eq:linearBCS_gap_eq}%
\end{equation}
In this case the order parameter is momentum independent, $\Delta
(\mathbf{p})=\Delta$, and Eq.~(\ref{eq:linearBCS_gap_eq}) reduces to the
equation for the critical temperature:%

\begin{equation}
1=\frac{\lambda}{2}\left[  \ln\frac{8\mu_{1}\exp(\gamma-2)}{\pi T_{BCS}}%
+\ln\frac{8\mu_{2}\exp(\gamma-2)}{\pi T_{BCS}}\right]  ,
\label{BCSlinear_gap_eq}%
\end{equation}
where $\gamma=0.5772$ is the Euler constant and we introduced a small
parameter
\begin{equation}
\label{lambda}\lambda=2p_{F}\left\vert a\right\vert /\pi\hbar\ll1.
\end{equation}

Equation (\ref{BCSlinear_gap_eq}) is obtained straightforwardly. First,
integrating Eq.~(\ref{eq:linearBCS_gap_eq}) over the angles one has%

\begin{align*}
1  &  =\frac{\lambda}{2}\int_{0}^{\infty}dx\left[  \tanh[(x^{2}-1)\mu
_{1}/2T_{c}]+\tanh[(x^{2}-1)\mu_{2}/2T_{c}]-2\right] \\
&  +\frac{\lambda}{2}\int_{0}^{\infty}dx\frac{\tanh[(x^{2}-1)\mu_{1}%
/2T_{c}]+\tanh[(x^{2}-1)\mu_{2}/2T_{c}]}{x^{2}-1},
\end{align*}
where $x=p/p_{F}$. For $\mu_{i}/T_{c}>>1$ the integrand of the first integral
is equal to $-4$ for $x<1$, and for $x>1$ it rapidly drops to $0$ in a narrow
interval of $x$, where $\left\vert x-1\right\vert \lesssim T_{c}/\mu<<1$. The
contribution of this interval can be neglected and, therefore, the first term
equals $-2 \lambda$. Then, after integrating the second term by parts, we obtain%

\[
1=\lambda\left\{  -2-\frac{1}{2}\int_{0}^{\infty}dx\left[  \frac{x\mu
_{1}/T_{c}}{\cosh^{2}[(x^{2}-1)\mu_{1}/2T_{c}]}+\frac{x\mu_{2}/T_{c}}%
{\cosh^{2}[(x^{2}-1)\mu_{2}/2T_{c}]}\right]  \ln\left\vert \frac{x-1}%
{x+1}\right\vert \right\}  .
\]
The final integration can easily be performed by using the fact that the
integrand is non-zero only in a narrow range of $x$, where $\left\vert
x-1\right\vert \lesssim T_{c}/\mu<<1$. We can therefore introduce a new
variable $y=x-1$ and extend the limits of integration over $y $ from $-\infty$
to $+\infty$. The equation then reads:%

\[
1=\lambda\left\{  -2-\frac{1}{2}\int_{-\infty}^{\infty}dy\left[  \frac{\mu
_{1}/T_{c}}{\cosh^{2}(y\mu_{1}/T_{c})}+\frac{\mu_{2}/T_{c}}{\cosh^{2}(y\mu
_{2}/T_{c})}\right]  \ln\frac{\left\vert y\right\vert }{2}\right\}  ,
\]
and performing the integration one arrives at Eq.~(\ref{BCSlinear_gap_eq}).
This equation gives the critical BCS temperature (cf. \cite{Caldas}):%

\begin{equation}
T_{BCS}=\frac{8}{\pi}\exp(\gamma-2)\sqrt{\mu_{1}\mu_{2}}\exp\left(  -\frac
{1}{\lambda}\right)  . \label{eq:TcBCS}%
\end{equation}

However, the linearized BCS gap equation (\ref{eq:linearBCS_gap_eq}) can only
be used for the calculation of the leading contribution to the critical
temperature, corresponding to the term $\sim\lambda^{-1}$ in the exponent of
Eq.~(\ref{eq:TcBCS}). Therefore, only the exponent in this equation is
correct. As was shown by Gorkov and Melik-Barkhudarov \cite{GMB}, the
preexponential factor in Eq.~(\ref{eq:TcBCS}) is determined by next-to-leading
order terms, which depend on many-body effects in the interparticle
interaction. These are the interactions between particles in a many-body
system through the polarization of the medium - virtual creation of
particle-hole pairs.

The importance of the many-body effects for the preexponential factor can be
understood as follows. After performing the integration over momenta, the gap
equation (\ref{eq:linear_ren_gap}) can be qualitatively written as%

\begin{equation}
1=\nu_{F}V_{\mathrm{eff}}\left[  \ln\frac{\mu}{T_{c}}+C\right]  ,
\label{eq:approx_eqTc}%
\end{equation}
where $\nu_{F}=m_{r}p_{F}/\pi^{2}\hbar^{3}$. In this formula, the large
logarithm $\ln\mu/T_{c}$ comes from the integration over momenta near the
Fermi surface, whereas the momenta far from the Fermi surface contribute to
the constant $C$ which is of the order of unity. We then write $\nu
_{F}V_{\mathrm{eff}}=-\lambda+a\lambda^{2}$, where the first term is the
direct interparticle interaction and we keep only the second order term in the
many-body part $\delta V$ of the effective interaction. It is now easy to see
that $\lambda\ln\mu/T_{c}\sim1$ and, therefore, the terms $a\lambda^{2}\ln
\mu/T_{c}$ and $\lambda C$ are of the same order of magnitude. As a result,
both terms have to be taken into account for the calculation of the
preexponential factor. Also, note that the contribution of the many-body part
of the interparticle interaction comes from momenta near the Fermi surface,
which are responsible for the large logarithm $\ln\mu/T_{c}\sim\lambda^{-1}$.
Hence, only the values of $\delta V$\ at the Fermi surface are important.

We now calculate the contribution of the many-body effects to the
preexponential factor for the critical temperature. They are usually called
Gorkov-Melik-Barkhudarov (GM) corrections. As it was argued above, in the weak
coupling limit the most important contributions to the effective interaction
are second order in $g$ (the role of high order terms will be discussed
later). In the considered case of a two-component Fermi gas with an $s$-wave
interaction, there is only the contribution shown in Fig. \ref{Fig1}, and the
corresponding analytical expression reads:%

\begin{equation}
\delta V(\mathbf{p},\mathbf{p}^{\prime})=-g^{2}\int\frac{d\mathbf{k}}%
{(2\pi)^{3}}\frac{f[\xi_{1}(\mathbf{k}+\mathbf{q}/2)]-f[\xi_{2}(\mathbf{k}%
-\mathbf{q}/2)]}{\xi_{1}(\mathbf{k}+\mathbf{q}/2)-\xi_{2}(\mathbf{k}%
-\mathbf{q}/2)}, \label{eq:V2}%
\end{equation}
where $\mathbf{q}=\mathbf{p}+\mathbf{p}^{\prime}$. In obtaining this
expression we used the zero-temperature distribution function $f[\xi
_{1,2}(p)]=\theta(-\xi_{1,2}(p))$, with $\theta(x)$ being the step function.
This is legitimate because the finite temperature corrections are proportional
to the ratio of the critical temperature to the chemical potential and,
therefore, are exponentially small. As can be seen from Eq.~(\ref{eq:V2}), the
effective interaction $\delta V(\mathbf{p},\mathbf{p}^{\prime})$ changes on
the momentum scale $p\sim p^{\prime}\sim p_{F}$.

We now solve Eq.~(\ref{eq:linear_ren_gap}). In this equation, the momentum
dependence of the order parameter originates only from the momentum dependence
of the many-body contribution to the interparticle interaction and, therefore,
contains an extra power of the small parameter $\lambda$. As a result, this
dependence can be ignored in the first integral on the right-hand side of Eq.
(\ref{eq:linear_ren_gap}), and we can simply replace there the order parameter
$\Delta(p^{\prime})$ by its value on the Fermi surface $\Delta(p_{F})$. This
does not affect the convergence of the integral at large momenta and, hence,
changes only the constant $C$ in Eq.~(\ref{eq:approx_eqTc}). The corresponding
modification, however, is proportional to the small parameter $\lambda$ and
can be neglected. In the second integral on the right-hand side of Eq.
(\ref{eq:linear_ren_gap}), as we have discussed earlier, only momenta
$p^{\prime}$ near the Fermi surface ($p^{\prime}\approx p_{F}$) are important,
and we can also put $p^{\prime}=p_{F}$ in $\Delta(p^{\prime})$ and $\delta
V(\mathbf{p},\mathbf{p}^{\prime})$. The gap equation then reads:%

\begin{align}
\Delta(\mathbf{p})  &  =-\frac{2\pi\hbar^{2}a}{m_{r}}\int\frac{d\mathbf{p}%
^{\prime}}{(2\pi\hbar)^{3}}\left[  \frac{\tanh[\xi_{1}(\mathbf{p}^{\prime
})/2T_{c}]+\tanh[\xi_{2}(\mathbf{p}^{\prime})/2T_{c}]}{2[\xi_{1}%
(\mathbf{p}^{\prime})+\xi_{2}(\mathbf{p}^{\prime})]}-\frac{2m_{r}}{p^{\prime
2}}\right]  \Delta(\mathbf{n}^{\prime}p_{F})\nonumber\\
&  -\int_{p<\Lambda p_{F}}\frac{d\mathbf{p}^{\prime}}{(2\pi\hbar)^{3}}\delta
V(\mathbf{p},\mathbf{n}^{\prime}p_{F})\frac{\tanh[\xi_{1}(\mathbf{p}^{\prime
})/2T_{c}]+\tanh[\xi_{2}(\mathbf{p}^{\prime})/2T_{c}]}{2[\xi_{1}%
(\mathbf{p}^{\prime})+\xi_{2}(\mathbf{p}^{\prime})]}\Delta(\mathbf{n}^{\prime
}p_{F}), \label{eq:linear_ren_gap1}%
\end{align}
where $\mathbf{n}^{\prime}$ is the unit vector in the direction of
$\mathbf{p}^{\prime}$ and we introduced an upper cut-off $\Lambda p_{F}$ with
$\Lambda\sim1$,\ for the purpose of convergence at large momenta. The exact
value of $\Lambda$ is not important because, as we mentioned above, the
contribution of large momenta to this integral has to be neglected. To derive
an equation for the critical temperature, we consider Eq.
(\ref{eq:linear_ren_gap1}) for $\mathbf{p}=\mathbf{n}p_{F}$ and average it
over the directions of $\mathbf{n}$. Taking into account that the order
parameter for the $s$-wave pairing can only depend on the absolute value of
the momentum, we obtain:%

\begin{align}
1  &  =-\frac{2\pi\hbar^{2}a}{m_{r}}\int\frac{d\mathbf{p}^{\prime}}{(2\pi
\hbar)^{3}}\left[  \frac{\tanh[\xi_{1}(\mathbf{p}^{\prime})/2T_{c}]+\tanh
[\xi_{2}(\mathbf{p}^{\prime})/2T_{c}]}{2[\xi_{1}(\mathbf{p}^{\prime})+\xi
_{2}(\mathbf{p}^{\prime})]}-\frac{2m_{r}}{p^{\prime2}}\right] \nonumber\\
&  -\overline{\delta V}\int_{p<\Lambda p_{F}}\frac{d\mathbf{p}^{\prime}}%
{(2\pi\hbar)^{3}}\frac{\tanh[\xi_{1}(\mathbf{p}^{\prime})/2T_{c}]+\tanh
[\xi_{2}(\mathbf{p}^{\prime})/2T_{c}]}{2[\xi_{1}(\mathbf{p}^{\prime})+\xi
_{2}(\mathbf{p}^{\prime})]}\nonumber\\
&  =\lambda\ln\frac{8\sqrt{\mu_{1}\mu_{2}}\exp(\gamma-2)}{\pi T_{c}}-\nu
_{F}\overline{\delta V}\ln\frac{8\sqrt{\mu_{1}\mu_{2}}\exp(\gamma+\Lambda
-2)}{\pi T_{c}}\approx(\lambda-\nu_{F}\overline{\delta V})\ln\frac{8\sqrt
{\mu_{1}\mu_{2}}\exp(\gamma-2)}{\pi T_{c}}, \label{eq:linear_ren_gap3}%
\end{align}
where%

\begin{equation}
\overline{\delta V}=\int\frac{d\mathbf{n}}{4\pi}\int\frac{d\mathbf{n}^{\prime
}}{4\pi}\delta V(\mathbf{n}p_{F},\mathbf{n}^{\prime}p_{F}) \label{deltaVint}%
\end{equation}
is the $s$-wave component of the many-body interaction. Using Eq.~(\ref{eq:V2}%
) and integrating over the angles in Eq.~(\ref{deltaVint}), we obtain%
\begin{equation}
\overline{\delta V}=\nu_{F}g^{2}\frac{1+\ln4}{3}[f(\kappa)+f(\kappa^{-1})],
\label{deltaVf}%
\end{equation}
where $\kappa=M/m$ and the function $f(\kappa)$ is given by%

\begin{equation}
f(\kappa)=-\frac{3(1+\kappa)}{4(1+\ln4)}\int_{0}^{1}dq\int_{0}^{1}%
pdp\ln\left\vert \frac{(p^{2}-1)(\kappa-1)+4q(p-q)}{(p^{2}-1)(\kappa
-1)-4q(p+q)}\right\vert . \label{eq:f}%
\end{equation}
A straightforward lengthy integration of Eq.~(\ref{eq:f}) yields%

\begin{align}
f(\kappa)  &  =-\frac{3}{4(1+\ln4)}(1+\kappa)\left[  -\frac{\kappa+1}{3\kappa
}\ln(2)+\frac{\kappa-1}{3\kappa}\ln\left\vert \kappa-1\right\vert \right.
\nonumber\\
&  \left.  +\frac{4}{3(\kappa-1)}-\left(  \frac{(\kappa+3)^{2}}{6(\kappa
-1)^{2}}+\frac{\kappa+2}{6\kappa}\right)  \ln\frac{\kappa+1}{2}\right]  .
\label{f_function}%
\end{align}
From Eqs.~(\ref{eq:linear_ren_gap3}) and (\ref{deltaVf}) we obtain the
following expression for the critical temperature:%
\[
T_{GM}=\frac{8}{\pi}e^{\gamma-2}\sqrt{\mu_{1}\mu_{2}}\exp[-(\lambda-\nu
_{F}\overline{\delta V})^{-1}]\approx T_{BCS}\exp(-\nu_{F}\overline{\delta
V}/\lambda^{2})
\]%
\begin{equation}
=\frac{e^{\gamma}}{\pi}\left(  \frac{2}{e}\right)  ^{7/3}\exp\left\{
-\frac{1+\ln4}{3}[f(\kappa)+f(\kappa^{-1})-1]\right\}  \sqrt{\mu_{1}\mu_{2}%
}\exp\left(  -\frac{1}{\lambda}\right)  . \label{eq:TcGMB}%
\end{equation}
It can be rewritten in the form%
\begin{equation}
T_{GM}=\frac{e^{\gamma}}{\pi}\left(  \frac{2}{e}\right)  ^{7/3}\mu_{1}%
F(\kappa)\exp\left(  -\frac{1}{\lambda}\right)  \equiv0.277\,\frac{p_{F}^{2}%
}{2M}F(\kappa)\exp(-\pi\hbar/2p_{F}|a|), \label{TcF}%
\end{equation}
where we expressed the Fermi energy of light atoms $\mu_{2}$ through the
heavy-atom Fermi energy $\mu_{1}=p_{F}^{2}/2M$, and the function $F(\kappa)$
is given by
\begin{equation}
F(\kappa)=\sqrt{\kappa}\exp\left\{  -\frac{1+\ln4}{3}[f(\kappa)+f(\kappa
^{-1})-1]\right\}  . \label{F_function}%
\end{equation}

\begin{figure}[ptb]
\begin{center}
\includegraphics[width=8cm]{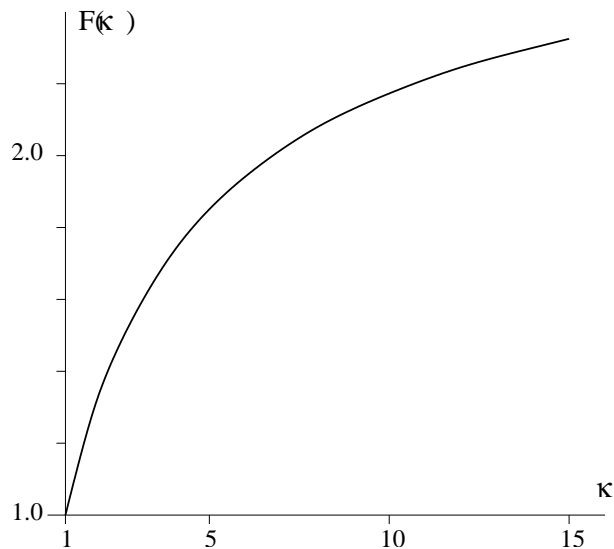}
\end{center}
\caption{The function $F(\kappa)$ in the preexponential factor of
Eq.~(\ref{TcF}).}%
\label{Fig2}%
\end{figure}

For equal masses one has $\kappa=M/m=1$, and Eqs.~(\ref{f_function}) and
(\ref{F_function}) give $f(1)=1/2$, $F(1)=1$. Then Eq.~(\ref{eq:TcGMB})
reproduces the original result of Ref. \cite{GMB}. The function $F(\kappa)$ is
shown in Fig. \ref{Fig2}. 
For a large mass ratio $\kappa=M/m\gg1$, this function tends to a
constant value, $F(\kappa\rightarrow\infty)=2^{4/3}e^{1/6}$. The critical
temperature is then given by%
\begin{equation}
T_{GM}=\frac{8e^{\gamma-2}}{\pi}2^{2/3}e^{-1/6}\mu_{1}\exp(-1/\lambda
)\equiv0.825\,\frac{p_{F}^{2}}{2M}\exp(-\pi\hbar/2p_{F}|a|),\quad\kappa\gg1.
\label{TcLargeratio}%
\end{equation}

Note that this result is quite different from the BCS critical temperature of
Eq.~(\ref{eq:TcBCS}). Aside from a constant of the order of unity, it contains
an extra small factor $1/\sqrt{\kappa}=\sqrt{m/M}\ll1$. Thus, for a fermionic
mixture with a large mass ratio of the components the second order
contribution significantly reduces the critical temperature compared to the
prediction of the simple BCS approach.

In Table 1 we show the critical temperature $T_{GM}$ following from Eq.
(\ref{TcF}) with $F(\kappa)$ given by Eq. (\ref{F_function}), for various
mixtures of fermionic atoms. The critical temperature is given in units of
$T_{GM}$ for a $^{6}$Li-$^{6}$Li mixture, and it is assumed that the quantity
$p_{F}^{2}\exp(-\pi\hbar/2p_{F}|a|)$ is the same for all mixtures. One clearly
sees that replacing one species in a mixture by a lighter one increases the
critical temperature, whereas the replacement with a heavier one decreases
$T_{GM}$.

\begin{table}[ptb]
\begin{center}%
\begin{tabular}
[c]{|c|c|c|c|c|}\hline
& $^{6}$Li & $^{40}$K & $^{87}$Sr & $^{171}$Yb\\\hline
$^{6}$Li & 1.000 & 0300 & 0.161 & 0.090\\\hline
$^{40}$K &  & 0.150 & 0.097 & 0.062\\\hline
$^{87}$Sr &  &  & 0.069 & 0.048\\\hline
$^{171}$Yb &  &  &  & 0.035\\\hline
\end{tabular}
\end{center}
\caption{Critical temperatures $T_{GM}$ for mixtures of various atomic species
as given by Eq. (\ref{TcF}). The scattering length $a$ and Fermi momentum
$p_{F}$ are assumed to be the same for all combinations, and $T_{GM}$ is in
units of the critical temperature for a $^{6}$Li-$^{6}$Li mixture.}%
\label{default}%
\end{table}

\section{Critical temperature. Higher order contributions}

Let us now consider the contribution of higher order ($\sim\lambda^{3}$)
many-body corrections. As can be seen from Eq. (\ref{eq:TcGMB}), these
corrections enter the exponent for the critical temperature being divided by
$\lambda^{2}$ and, therefore, the corresponding term is $\lambda Q(M/m)$,
where $Q$ is a function of the mass ratio. For moderate values of $M/m$, the
function $Q$ is of the order of unity, and, therefore, the corresponding
corrections can be neglected. However, as we will see later, for a large mass
ratio, the function $Q$ becomes proportional to $M/m$, and the related
corrections in the exponent are $\sim k_{F}aM/m$. The applicability of the
perturbation theory requires $(k_{F}a)^{2}M/m\ll1$ (see Eq.
(\ref{small_parameter}) below), but the quantity $k_{F}aM/m$ should not
necessarily be small for $M/m\gg1$. As a result, the third-order many-body
corrections proportional to $M/m$ have to be taken into account.

For calculating the terms of the order of $\lambda^{3}$, we have to modify the
gap equation (\ref{eq:bcsgap}) in order to include the difference between
particles and quasiparticles, or single-particle excitations \cite{foot}. In
the considered case of a two-component Fermi gas, the quasiparticles are
characterized by the effective masses $M^{\ast}$ and $m^{\ast}$ and by the
$Z$-factors $Z_{M}$ and $Z_{m}$. The $Z$-factors are related to the amplitude
of creating a quasiparticle by adding an extra particle to the system (see
\cite{LL9} for rigorous definitions and details). For a given (effective)
interaction $V_{\mathrm{eff}}$ between heavy and light fermions, the
interaction between the corresponding quasiparticles is simply equal to
$Z_{M}Z_{m}V_{\mathrm{eff}}$. It is important for our approach that the ratios
$M^{\ast}/M$, $m^{\ast}/m$, and the constants $Z_{M}$, $Z_{m}$ differ from
unity only by a small amount proportional $\lambda$.

We can now extend the analysis of the gap equation to higher order terms. As
follows from the previous discussions, the qualitative form of the gap
equation can be written as%
\begin{equation}
1=\nu_{F}^{\ast}Z_{M}Z_{m}V_{\mathrm{eff}}\left[  \ln\frac{\mu}{T_{c}%
}+C\right]  , \label{gap3}%
\end{equation}
where $\nu_{F}^{\ast}=m_{r}^{\ast}k_{F}/\pi^{2}\hbar^{2}$ and the reduced mass
$m_{r}^{\ast}$ is determined by the effective masses $M^{*}$ and $m^{*}$. The
effective interaction $V_{\mathrm{eff}}=g+\delta V+\delta V^{(3)}$ now
includes also the third-order many-body contribution $\delta V$ $^{(3)}$. We
are interested in the terms of the order of $\lambda^{2}$ and $\lambda^{3}%
\ln\mu/T_{c}$ on the right-hand side of Eq. (\ref{gap3}), where the
$\lambda^{3}\ln\mu/T_{c}$ contributions come from the integration over momenta
near the Fermi surface, whereas the momenta far from the Fermi surface result
in $\lambda^{2}$ contributions. The term of the order $\lambda^{2}$ in Eq.
(\ref{gap3}) comes only from the second-order term $\delta V$ in the effective
interaction. For a large mass ratio this term contains only $\ln(M/m)$ for
large mass ratio (see. Eqs. (\ref{deltaVf}) and \ref{f_function}) and,
therefore, can be neglected. The term $\lambda^{3}\ln\mu/T_{c}$ results from
the third-order term $\delta V^{(3)}$ in the effective interaction (with
$\nu_{F}^{\ast}Z_{M}Z_{m}\rightarrow\nu_{F}$) and from the difference between
particles and quasiparticles, $(\nu_{F}^{\ast}Z_{M}Z_{m}-\nu_{F})\sim
\lambda^{2}$, multiplied by the first order term $g$ in the effective
interaction. As a result, up to terms of the order of $\lambda^{2}$, we can
write the linearized renormalized gap equation as%
\begin{align}
\Delta(\mathbf{p})  &  =-\frac{2\pi\hbar^{2}a}{m_{r}}\frac{m_{r}^{\ast}}%
{m_{r}}Z_{M}Z_{m}\int\frac{d\mathbf{p}^{\prime}}{(2\pi\hbar)^{3}}\left[
\frac{\tanh[\xi_{1}(\mathbf{p}^{\prime})/2T_{c}]+\tanh[\xi_{2}(\mathbf{p}%
^{\prime})/2T_{c}]}{2[\xi_{1}(\mathbf{p}^{\prime})+\xi_{2}(\mathbf{p}^{\prime
})]}-\frac{2m_{r}}{p^{\prime2}}\right]  \Delta(\mathbf{n}^{\prime}%
p_{F})\nonumber\\
&  -\int_{p<\Lambda p_{F}}\frac{d\mathbf{p}^{\prime}}{(2\pi\hbar)^{3}}\left[
\delta V(\mathbf{p},\mathbf{n}^{\prime}p_{F})+\delta V^{(3)}(\mathbf{p}%
,\mathbf{n}^{\prime}p_{F})\right]  \frac{\tanh[\xi_{1}(\mathbf{p}^{\prime
})/2T_{c}]+\tanh[\xi_{2}(\mathbf{p}^{\prime})/2T_{c}]}{2[\xi_{1}%
(\mathbf{p}^{\prime})+\xi_{2}(\mathbf{p}^{\prime})]}\Delta(\mathbf{n}^{\prime
}p_{F}). \label{lingap3}%
\end{align}
As in Eq. (\ref{eq:linear_ren_gap1}), we introduce an upper cut-off $\Lambda
p_{F}$ for the purpose of convergence of integrals at large momenta. Eq.
(\ref{lingap3}) can be solved in the same way as Eq. (\ref{eq:linear_ren_gap1}%
) and we obtain%
\begin{align}
T_{c}  &  =\frac{8}{\pi}e^{\gamma-2}\sqrt{\mu_{1}\mu_{2}}\exp\left\{  -\left[
\frac{m_{r}^{\ast}}{m_{r}}Z_{M}Z_{m}\lambda-\nu_{F}\left(  \overline{\delta
V}+\overline{\delta V^{(3)}}\right)  \right]  ^{-1}\right\} \nonumber\\
&  \approx T_{BCS}\exp\left\{  -\frac{1}{\lambda^{2}}\left[  \left(
\frac{m_{r}^{\ast}}{m_{r}}Z_{M}Z_{m}-1\right)  \lambda+\nu_{F}\left(
\overline{\delta V}+\overline{\delta V^{(3)}}\right)  \right]  \right\}
\nonumber\\
&  =T_{cGM}\exp\left\{  -\frac{1}{\lambda^{2}}\left[  \left(  \frac
{m_{r}^{\ast}}{m_{r}}Z_{M}Z_{m}-1\right)  \lambda+\nu_{F}\overline{\delta
V^{(3)}}\right]  \right\}  . \label{Tc3}%
\end{align}
Note that only the contributions that are linear in $M/m$ for $M/m\gg1$ should
be kept in $m_{r}^{\ast}/m_{r}$, $Z_{M}$, $Z_{m}$, and $\overline{\delta
V^{(3)}}$.

\begin{figure}[ptb]
\begin{center}
\includegraphics[width=15cm]{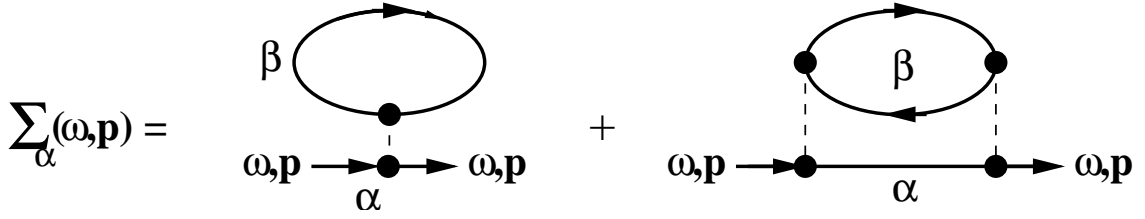}
\end{center}
\caption{The first and second order diagrams to the self-energy $\Sigma
_{\alpha}(\omega,p)$. Note that we present here only irreducible diagrams,
and, therefore, the diagram of the second order containing the first order
self-energy insertion to the Green function of the $\beta$-fermion in the
first diagram is omitted.}%
\label{FigX}%
\end{figure}

The effective masses $M^{\ast}$, $m^{\ast}$ and the constants $Z_{M}$, $Z_{m}$
can be obtained from the derivatives of the corresponding self-energies
$\Sigma_{M}(\omega,p)$ and $\Sigma_{m}(\omega,p)$ with respect to the
frequency $\omega$ and momentum $p$, evaluated at $\omega=0$ and $p=p_{F}$:%
\begin{align}
Z_{M(m)}  &  =\left(  1-\left.  \frac{\partial\Sigma_{M(m)}(\omega
,p)}{\partial\omega}\right\vert _{\omega=0,p=p_{F}}\right)  ^{-1},\label{Z}\\
M^{\ast}/M  &  =Z_{M}^{-1}\left(  1+\frac{M}{p_{F}}\left.  \frac
{\partial\Sigma_{M}(\omega,p)}{\partial p}\right\vert _{\omega=0,p=p_{F}%
}\right)  ^{-1},\label{Meff}\\
m^{\ast}/m  &  =Z_{m}^{-1}\left(  1+\frac{m}{p_{F}}\left.  \frac
{\partial\Sigma_{m}(\omega,p)}{\partial p}\right\vert _{\omega=0,p=p_{F}%
}\right)  ^{-1}. \label{meff}%
\end{align}
The diagrams for the self energies $\Sigma_{M}$ and $\Sigma_{m}$ up to the
second order in $g$ are shown in Fig. \ref{FigX}. 
The corresponding analytical expressions read%
\begin{align}
\Sigma_{\alpha}(\omega,p)  &  =\frac{2\pi\hbar^{2}a}{m_{r}}n_{\beta}+\left(
\frac{2\pi\hbar^{2}a}{m_{r}}\right)  ^{2}\int\frac{d\omega_{2}}{2\pi}%
\frac{d\mathbf{p}_{2}}{(2\pi\hbar)^{3}}G_{\beta}(\omega_{2},\mathbf{p}%
_{2})\nonumber\\
&  \times\int\frac{d\omega_{1}}{2\pi}\frac{d\mathbf{p}_{1}}{(2\pi\hbar)^{3}%
}\left[  G_{\beta}(\omega_{1}+\omega_{2},\mathbf{p}_{2}+\mathbf{p}%
_{1})G_{\alpha}(-\omega_{1}+\omega,-\mathbf{p}_{1}+\mathbf{p})-G_{\beta}%
^{(0)}(\omega_{1},\mathbf{p}_{1})G_{\alpha}^{(0)}(-\omega_{1},-\mathbf{p}%
_{1})\right]  , \label{sigma}%
\end{align}
where $\alpha=M,\,\beta=m$ or $\alpha=m,\,\beta=M$, and the Green functions
are given by
\[
G_{\alpha(\beta)}(\omega,\mathbf{p})=\frac{1}{\omega-\xi_{\alpha(\beta
)}(\mathbf{p})+i\delta\mathrm{sign}[\xi_{\alpha(\beta)}(\mathbf{p})]},
\]%
\[
G_{\alpha(\beta)}^{(0)}(\omega,\mathbf{p})=\frac{1}{\omega-p^{2}%
/2m_{\alpha(\beta)}+i\delta}%
\]
with $\delta=+0$. The divergent integral in the second-order contribution is
renormalized in a standard way by replacing the coupling constant $g$ with the
scattering amplitude $a$ and subtracting the product $G_{\beta}^{(0)}%
(\omega_{1},\mathbf{p}_{1})G_{\alpha}^{(0)}(-\omega_{1},-\mathbf{p}_{1})$ of
the two Green functions in vacuum ($\mu_{1,2}=0$), which corresponds to the
second order Born contribution to the scattering amplitude, from the
integrand. After integrating over the frequencies $\omega_{1}$ and $\omega
_{2}$ in Eq. (\ref{sigma}), we obtain%
\begin{align}
\Sigma_{\alpha}(\omega,p)  &  =\frac{2\pi\hbar^{2}a}{m_{r}}n+\left(
\frac{2\pi\hbar^{2}a}{m_{r}}\right)  ^{2}\int\frac{d\mathbf{p}_{2}}{(2\pi
\hbar)^{3}}\frac{d\mathbf{p}_{1}}{(2\pi\hbar)^{3}}\left\{  \frac
{(1-f[\xi_{\beta}(\mathbf{p}_{2})])f[\xi_{\alpha}(-\mathbf{p}_{1}%
+\mathbf{p})]f[\xi_{\beta}(\mathbf{p}_{2}+\mathbf{p}_{1})]}{\omega-\xi
_{\alpha}(-\mathbf{p}_{1}+\mathbf{p})-\xi_{\beta}(\mathbf{p}_{2}%
+\mathbf{p}_{1})+\xi_{\beta}(\mathbf{p}_{2})-i\delta}\right. \nonumber\\
&  +\left.  \frac{f[\xi_{\beta}(\mathbf{p}_{2})](1-f[\xi_{\alpha}%
(-\mathbf{p}_{1}+\mathbf{p})])(1-f[\xi_{\beta}(\mathbf{p}_{2}+\mathbf{p}%
_{1})])}{\omega-\xi_{\alpha}(-\mathbf{p}_{1}+\mathbf{p})-\xi_{\beta
}(\mathbf{p}_{2}+\mathbf{p}_{1})+\xi_{\beta}(\mathbf{p}_{2})+i\delta}%
+\frac{f[\xi_{\beta}(\mathbf{p}_{2})]}{p_{1}^{2}/2m_{r}-i\delta}\right\}  .
\label{sigmaP}%
\end{align}
As we discussed above, for a large $M/m$ only the leading contributions that
are linear in $M/m$ should be kept, and lengthy calculations with the use of
Eqs. (\ref{Z})-(\ref{sigmaP}) give%
\begin{align}
Z_{M}  &  =1,\label{ZM}\\
\frac{M^{\ast}}{M}  &  =1+\frac{1}{5}\left(  2\ln2-1\right)  \left(
\frac{ap_{F}}{\pi\hbar}\right)  ^{2}\frac{M}{m},\label{Meffres}\\
Z_{m}  &  =1-\frac{1}{3}(1+2\ln2)\left(  \frac{ap_{F}}{\pi\hbar}\right)
^{2}\frac{M}{m},\label{Zm}\\
\frac{m^{\ast}}{m}  &  =1+\frac{1}{3}(1+2\ln2)\left(  \frac{ap_{F}}{\pi\hbar
}\right)  ^{2}\frac{M}{m}. \label{meffres}%
\end{align}

\begin{figure}[ptb]
\begin{center}
\includegraphics[width=14cm]{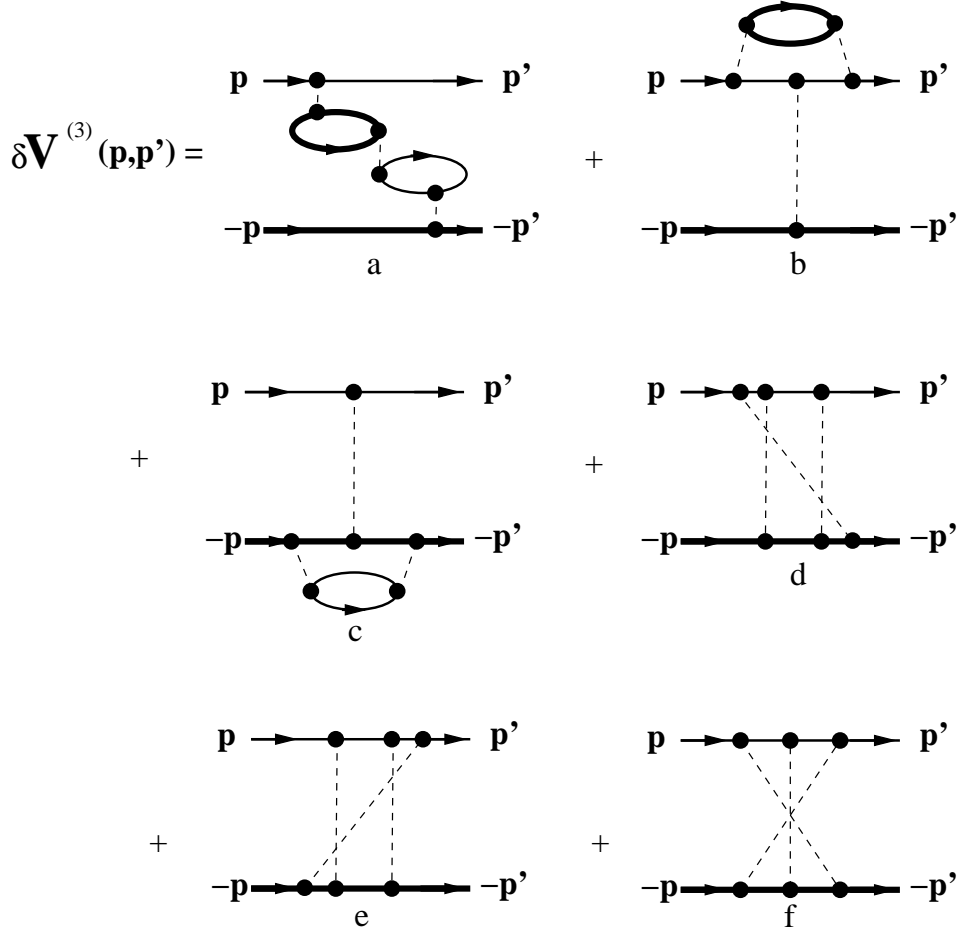}
\end{center}
\caption{The third-order contributions to the effective interaction between
heavy (thick line) and light (thin line) fermions. The dashed line corresponds
to the coupling constant $g$.}%
\label{FigY}%
\end{figure}

The diagrams for third-order contributions to the effective interaction
$V_{\mathrm{eff}}(\mathbf{p},\mathbf{p}^{\prime})$ are shown in Fig.
\ref{FigY}, where we omit the diagrams that can be obtained by inserting the
first order self-energy blocks (the first diagram in Fig. \ref{FigX}) into the
internal lines of the second-order diagram from Fig. \ref{Fig1} . These
self-energy contributions (the first term in Eq. (\ref{sigma})) simply shift
the chemical potentials. It turns out that only diagrams a, b, and c could
contain terms linear in $M/m$, whereas the rest of the diagrams are
proportional to $\ln(M/m)$. The divergencies at large momenta in diagrams d
and f can be removed by renormalizing the coupling constant $g$ in the
second-order diagram in Fig. \ref{Fig1}. Analytical expressions for the
diagrams a,b, and c are the following:
\[
\delta V_{a}^{(3)}(\mathbf{p},\mathbf{p}^{\prime})=g^{3}\int\frac{d\omega
_{1}d\mathbf{p}_{1}}{(2\pi)^{4}\hbar^{3}}G_{M}(\omega_{1},\mathbf{q}%
+\mathbf{p}_{1})G_{M}(\omega_{1},\mathbf{p}_{1})\int\frac{d\omega
_{2}d\mathbf{p}_{2}}{(2\pi)^{4}\hbar^{3}}G_{m}(\omega_{2},\mathbf{q}%
+\mathbf{p}_{2})G_{m}(\omega_{2},\mathbf{p}_{2}),
\]

\[
\delta V_{b}^{(3)}(\mathbf{p},\mathbf{p}^{\prime})=g^{3}\int\frac{d\omega
_{1}d\mathbf{p}_{1}}{(2\pi)^{4}\hbar^{3}}G_{M}(\omega_{1},\mathbf{p}%
+\mathbf{p}_{1})G_{M}(\omega_{1},\mathbf{p}^{\prime}+\mathbf{p}_{1})\int
\frac{d\omega_{2}d\mathbf{p}_{2}}{(2\pi)^{4}\hbar^{3}}G_{m}(\omega
_{2},\mathbf{p}_{2})G_{m}(\omega_{1}+\omega_{2},\mathbf{p}_{1}+\mathbf{p}%
_{2}),
\]

\[
\delta V_{c}^{(3)}(\mathbf{p},\mathbf{p}^{\prime})=g^{3}\int\frac{d\omega
_{1}d\mathbf{p}_{1}}{(2\pi)^{4}\hbar^{3}}G_{m}(\omega_{1},\mathbf{p}%
+\mathbf{p}_{1})G_{m}(\omega_{1},\mathbf{p}^{\prime}+\mathbf{p}_{1})\int
\frac{d\omega_{2}d\mathbf{p}_{2}}{(2\pi)^{4}\hbar^{3}}G_{M}(\omega
_{2},\mathbf{p}_{2})G_{M}(\omega_{1}+\omega_{2},\mathbf{p}_{1}+\mathbf{p}%
_{2}),
\]
where $\mathbf{q}=\mathbf{p}-\mathbf{p}^{\prime}$. The integration over
frequencies $\omega_{1}$ and $\omega_{2}$ in the above expressions is
straightforward and gives%
\begin{align*}
\delta V_{a}^{(3)}(\mathbf{p},\mathbf{p}^{\prime})  &  =g^{3}\nu_{M}\nu
_{m}\int\frac{d\mathbf{p}_{1}}{(2\pi\hbar)^{3}}\frac{f[\xi_{1}(\mathbf{p}%
_{1}+\mathbf{q})]-f[\xi_{1}(\mathbf{p}_{1})]}{\xi_{1}(\mathbf{p}%
_{1}+\mathbf{q})-\xi_{1}(\mathbf{p}_{1})}\int\frac{d\mathbf{p}_{2}}{(2\pi
\hbar)^{3}}\frac{f[\xi_{2}(\mathbf{p}_{2}+\mathbf{q})]-f[\xi_{2}%
(\mathbf{p}_{2})]}{\xi_{2}(\mathbf{p}_{2}+\mathbf{q})-\xi_{2}(\mathbf{p}_{2}%
)}\\
&  =g^{3}\nu_{M}\nu_{m}\frac{1}{4}\left[  1+\frac{p_{F}}{q}\left(
1-\frac{q^{2}}{4p_{F}^{2}}\right)  \ln\frac{2p_{F}+q}{\left\vert
2p_{F}-q\right\vert }\right]  ^{2},
\end{align*}
where $\nu_{M}=Mp_{F}/2\pi^{2}\hbar^{3}$ and $\nu_{m}=mp_{F}/2\pi^{2}\hbar
^{3}$ are the densities of states at the Fermi level for heavy and light
fermions, respectively. For the other two contributions we obtain:
\begin{align*}
\delta V_{b}^{(3)}(\mathbf{p},\mathbf{p}^{\prime})  &  =g^{3}\int
\frac{d\mathbf{p}_{1}}{(2\pi\hbar)^{3}}\int_{0}^{\infty}dsA_{M}(s,\mathbf{p}%
_{1})\left[  \frac{f[\xi_{2}(\mathbf{p}_{1}+\mathbf{p})]}{\xi_{2}%
(\mathbf{p}_{1}+\mathbf{p})-\xi_{2}(\mathbf{p}_{1}+\mathbf{p}^{\prime
})+i\delta}\,\,\,\,\frac{1}{\xi_{2}(\mathbf{p}_{1}+\mathbf{p})-s}\right. \\
&  \left.  +\frac{1-f[\xi_{2}(\mathbf{p}_{1}+\mathbf{p})]}{\xi_{2}%
(\mathbf{p}_{1}+\mathbf{p})-\xi_{2}(\mathbf{p}_{1}+\mathbf{p}^{\prime
})-i\delta}\,\,\,\,\frac{1}{\xi_{2}(\mathbf{p}_{1}+\mathbf{p})+s}%
+(\mathbf{p}\leftrightarrow\mathbf{p}^{\prime})\right]
\end{align*}
with%
\begin{equation}
A_{M}(s,\mathbf{p})=\frac{Mp_{F}^{2}}{8\pi^{2}\hbar^{3}p}\left\{
\begin{array}
[c]{l}%
\displaystyle{1-\left(  \frac{Ms}{pp_{F}}-\frac{p}{2p_{F}}\right)  ^{2}%
,\quad\frac{p}{2M}\left\vert p-2p_{F}\right\vert \leq s\leq\frac{p}%
{2M}(p+2p_{F})}\\
\displaystyle{\frac{2Ms}{p_{F}^{2}},\quad0\leq s\leq\frac{p}{2M}(2p_{F}%
-p)};\,\,\,\,\,p\leq2p_{F},
\end{array}
\right.  , \label{A1}%
\end{equation}
and zero otherwise. A similar expression is obtained for $\delta V_{c}%
^{(3)}(\mathbf{p},\mathbf{p}^{\prime})$, with the replacements $\xi
_{2}\rightarrow\xi_{1}$ and $A_{M}(s,\mathbf{p})\rightarrow A_{m}%
(s,\mathbf{p})$.

The corresponding contributions to the $s$-wave scattering channel can be
obtained by averaging over the directions of the momenta $\mathbf{p}$ and
$\mathbf{p}^{\prime}$:%
\[
\overline{\delta V_{j}^{(3)}}=\int\frac{d\widehat{\mathbf{p}}}{4\pi}\int
\frac{d\widehat{\mathbf{p}}^{\prime}}{4\pi}\delta V_{j}^{(3)}(\mathbf{p}%
,\mathbf{p}^{\prime}),\quad j=a,b,c.
\]
In the limit of $M/m\gg1$, the leading terms in these contributions are%
\begin{align}
\overline{\delta V_{a}^{(3)}}  &  =g\frac{2+7\zeta(3)}{16}\left(  \frac
{ap_{F}}{\pi\hbar}\right)  ^{2}\frac{M}{m},\label{Va}\\
\overline{\delta V_{b}^{(3)}}  &  =0,\label{Vb}\\
\overline{\delta V_{c}^{(3)}}  &  =-g\frac{1+4(2+3\ln2)\ln2}{18}\left(
\frac{ap_{F}}{\pi\hbar}\right)  ^{2}\frac{M}{m}, \label{Vc}%
\end{align}
where $\zeta(x)$ is the Riemann zeta-functions ($\zeta(3)=1.202$). As a
result, the quantity $\overline{\delta V^{(3)}}$ in Eq. (\ref{Tc3}) is
\begin{equation}
\overline{\delta V^{(3)}}=\overline{\delta V_{a}^{(3)}}+\overline{\delta
V_{b}^{(3)}}+\overline{\delta V_{c}^{(3)}}. \label{V3}%
\end{equation}

Note that the validity of the perturbation theory requires the quantity
$\overline{\delta V^{(3)}}$ be smaller than the coupling constant $g$. This
leads to the condition
\begin{equation}
(ap_{F}/\hbar)^{2}M/m\ll1. \label{small_parameter}%
\end{equation}
Thus, the actual small parameter of the theory in the limit of a large mass
ratio $M/m$ is $(p_{F}|a|/\hbar)\sqrt{M/m}$.

After substituting Eqs. (\ref{ZM})-(\ref{meffres}) and (\ref{V3}) into Eq.
(\ref{Tc3}) we find the critical temperature in the limit of $M/m\gg1$:
\begin{equation}
T_{c}=\frac{8e^{\gamma-2}}{\pi}2^{2/3}e^{-1/6}\frac{p_{F}^{2}}{2M}\exp\left(
-\frac{\pi\hbar}{2\left\vert a\right\vert p_{F}}-0.034\frac{\left\vert
a\right\vert p_{F}}{\pi\hbar}\frac{M}{m}\right)  =T_{GM}\exp\{-0.011(p_{F}%
|a|/\hbar)M/m\}. \label{Treal}%
\end{equation}
Compared to the transition temperature in the GM approach, Eq.(\ref{Treal})
contains an extra exponential factor which, in principle, can be large.
However, this requires a very high mass ratio $M/m$. The extra term in the
exponent of Eq.(\ref{Treal}) can be written as $0.01\sqrt{M/m}\times
(p_{F}|a|/\hbar)\sqrt{M/m}$ and, since the second multiple in this expression
is small, one should have the mass ratio at least of the order of thousands in
order to get a noticeable change of $T_{c}$ compared to the GM result. In this
case Eq.(\ref{small_parameter}) shows that the parameter $\lambda
=2p_{F}|a|/\pi\hbar$ should be very small and, hence, the transition
temperature itself is vanishingly low.

We thus see that for reasonable values of $p_{F}|a|/\hbar$ satisfying
Eq.(\ref{small_parameter}), let say $p_{F}|a|/\hbar\sim0.1$ and $M/m<100$, the
higher order contributions do not really change the GM result for the
transition temperature.

At the same time, our analysis shows that for $M/m\gg1$ the conventional
weakly interacting regime requires much lower values of $\lambda=2p_{F}%
|a|/\pi\hbar$ than in the case of equal masses and the small parameter of the
perturbation theory is given by Eq.(\ref{small_parameter}) In the next section
we discuss the physical origin of this parameter.

\section{Small parameter of the theory}

There are several conditions that allow one to develop a perturbation theory
for a many-body fermionic system on the basis of Hamiltonian (\ref{H}). First
of all, this is the condition of the weakly interacting regime, which assumes
that the amplitude $a$ of the interspecies interaction is much smaller than
the mean separation between particles. The latter is of the order of
$\hbar/p_{F}$, and we immediately have the inequality
\begin{equation}
\label{ka}p_{F}|a|/\hbar\ll1.
\end{equation}
At the same time, inequality (\ref{ka}) allows one to use the binary approach
for the interparticle interaction. Then, assuming a short-range character of
the interatomic potential, one can consider the interaction between particles
as contact and write the interaction part of the Hamiltonian as $g\int
d\mathbf{r} \widehat{\psi}_{1}^{+}(\mathbf{r})\widehat{\psi}_{1}%
(\mathbf{r})\widehat{\psi}_{2}^{+} (\mathbf{r})\widehat{\psi}_{2}(\mathbf{r})$.

In the weakly interacting regime only fermions near the Fermi surface
participate in the response of the system to external perturbations.
Therefore, there is another condition that is needed for constructing the
perturbation theory. Namely, we have to assume that for both light and heavy
fermions the density of states near the Fermi surface is not strongly
distorted by the interactions. This is certainly the case if both Fermi
energies, $p_{F}^{2}/2m$ and $p_{F}^{2}/2M$, greatly exceed the mean-field
interaction $ng$. For $M\sim m$ this condition is equivalent to inequality
(\ref{ka}). In contrast, for $M\gg m$ Eq.~(\ref{ka}) only guarantees that the
Fermi energy of light fermions is $p_{F}^{2}/2m\gg ng$, whereas the condition
$p_{F}^{2}/2M\gg ng$ leads to the inequality $(p_{F}|a|/\hbar)M/m\ll1$. This
mean-field condition, however, is far too strong because at the mean-field
level, the interaction shifts uniformly all energy states and, hence, results
only in the change of the chemical potential. Actually, the
interaction-induced modification of the density of states is determined by the
momentum and frequency dependence of the fermionic self-energy (see
Eqs.~(\ref{Z})-(\ref{meff})) and appears in the second order of the
perturbative expansion in $g$ (the second diagram in Fig. \ref{FigX}). The
corresponding contribution describes the process in which a heavy fermion
pushes a light one out of the Fermi sphere and then, interacting once more
with this light fermion, puts it back to the initial state. Due to the Pauli
principle, the momenta of both light and heavy fermions in the intermediate
state should be larger than the Fermi momentum. As a result, for the initial
heavy-fermion state close to the Fermi surface, the most important
intermediate states will be those with momenta close to the Fermi momentum.
Therefore, the resulting contribution should be proportional to the product of
the densities of states of heavy and light fermions at the Fermi surface, and
the relative change of both densities of states is controlled by the parameter
$g^{2}\nu_{M}\nu_{m}\sim(p_{F}a/\hbar)^{2}M/m$. Thus, this parameter should be
small, i.e. we arrive at Eq.~(\ref{small_parameter}):
\[
(p_{F}a/\hbar)^{2}M/m\ll1.
\]

A complementary physical argument on support of this small parameter comes
from the consideration of the effective interaction between a light and a
heavy fermion in the medium. For example, the process described by the diagram
in Fig. 4a can be viewed in the following way. Incoming heavy and light
fermions interact with fermions inside the filled Fermi spheres and transfer
them to the states above the Fermi surfaces. Then the transferred heavy and
light fermions interact with each other and return to their initial states.
The important point is that the intermediate state of this process contains
excitations (particle-hole pairs) near the Fermi surface of the filled Fermi
sphere of heavy fermions. Therefore, the corresponding contribution to the
effective interaction is $g_{\mathrm{eff}}\sim g^{3}\nu_{m}\nu_{M}$, where the
densities of states of heavy and light fermions near the Fermi surface are
$\nu_{M}=Mp_{F}/2\pi^{2}\hbar^{3}$ and $\nu_{m}=mp_{F}/2\pi^{2}\hbar^{3}$,
respectively. This leads to $g_{\mathrm{eff}}\sim g(p_{F}a/\hbar)^{2}M/m$.
Comparing it with the direct interaction $g$ and requiring the inequality
$|g_{\mathrm{eff}}|\ll|g|$ which allows one to use a perturbation theory, we
again obtain a small parameter of the theory $(p_{F}a/\hbar)^{2}M/m\ll1$.

Let us now understand in which physical quantities the parameter
(\ref{small_parameter}) enters directly. In the limit of $M/m\gg1$ heavy
fermions occupy the energy interval $p_{F}^{2}/2M$ which is much narrower than
the energy interval $p_{F}^{2}/2m$ occupied by light fermions. However, the
heavy-fermion density of states is much larger: $\nu_{M}=Mp_{F}/(2\pi^{2}%
\hbar^{3})\gg\nu_{m}=mp_{F}/(2\pi^{2}\hbar^{3})$. The high density of states
of heavy fermions manifests itself in any quantity characterized by processes
where, for the heavy fermions, only the states near the Fermi surface are
important. This is the case for the effective masses of atoms, critical
temperature, and (see the next section) for the zero-temperature order
parameter $\Delta_{0}$. If, however, all states of the heavy fermions are
important, then the peak at $E\sim p_{F}^{2}/2M$ in the energy distribution of
heavy fermions is integrated out, and the result does not contain the
parameter (\ref{small_parameter}). This is exactly what is happening in the
calculation of the second order correction to the energy of the system, which
involves the sum over all energy states.

\begin{figure}[ptb]
\begin{center}
\includegraphics[width=8cm]{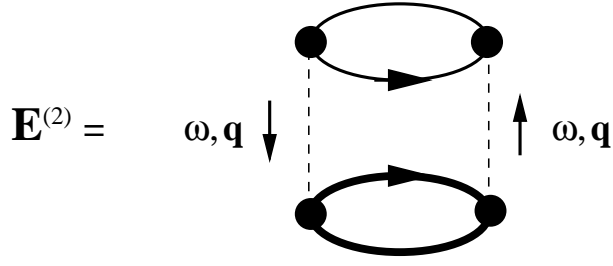}
\end{center}
\caption{The second order contribution to the energy of the system.}%
\label{FigE}%
\end{figure}

The second order contribution to the energy is shown diagrammatically in Fig.
\ref{FigE}, and the corresponding analytical expression reads%
\begin{equation}
E^{(2)}=g^{2}\int\frac{d\omega}{2\pi}\frac{d\mathbf{q}}{(2\pi\hbar)^{3}}%
\Pi_{m}(\omega,q)\Pi_{M}(-\omega,q), \label{energy2}%
\end{equation}
where%
\begin{align*}
\Pi_{\alpha}(\omega,q)  &  =\int\frac{d\omega_{1}}{2\pi}\frac{d\mathbf{p}_{1}%
}{(2\pi\hbar)^{3}}G_{\alpha}(\omega+\omega_{1},\mathbf{q}+\mathbf{p}%
_{1})G_{\alpha}(\omega_{1},\mathbf{p}_{1})\\
&  =-\int\frac{d\mathbf{p}_{1}}{(2\pi\hbar)^{3}}\frac{f[\xi_{\alpha
}(\mathbf{p}_{1}+\mathbf{q})]-f[\xi_{\alpha}(\mathbf{p}_{1})]}{\omega
-(\xi_{\alpha}(\mathbf{p}_{1}+\mathbf{q})-\xi_{\alpha}(\mathbf{p}%
_{1}))+i\delta(\mathrm{sign}[\xi_{\alpha}(\mathbf{p}_{1}+\mathbf{q}%
)]-\mathrm{sign}[\xi_{\alpha}(\mathbf{p}_{1})])}%
\end{align*}
is the polarization operator (the bubble in the diagrammatic language) for
$\alpha$-fermions ($\alpha=M$ or $m$). As it can be seen, the integration over
$\omega$ in Eq.~(\ref{energy2}) results in an integral that diverges at large
$q$. This divergence, however, can be eliminated by subtracting the
second-order Born contribution to the interparticle scattering amplitude
multiplied by the densities of fermions. This corresponds to the
renormalization of the coupling constant $g$ in the first order (mean-field)
contribution to the energy $E^{(1)}=gn^{2}$ (see \cite{LL9} for more details).
The resulting expression then coincides with equation (6.12) in \cite{LL9}.

After using the spectral representation for the polarization operator
$\Pi_{\alpha}(\omega,p)$:
\[
\Pi_{\alpha}(\omega,q)=\int_{0}^{\infty}dsA_{\alpha}(s,q)\left[  \frac
{1}{\omega-s+i\delta}-\frac{1}{\omega+s-i\delta}\right]  ,
\]
with the function $A_{\alpha}(s,p)$ from Eq. (\ref{A1}), equation
(\ref{energy2}) can be rewritten in the form%
\begin{equation}
E^{(2)}=\frac{1}{2}g^{2}\int\frac{d\mathbf{q}}{(2\pi\hbar)^{3}}\left[
\int_{0}^{\infty}ds_{1}\int_{0}^{\infty}ds_{2}\frac{A_{M}(s_{1},q)A_{m}%
(s_{2},q)}{s_{1}+s_{2}}-\frac{2m_{r}}{q^{2}}n^{2}\right]  , \label{E2A}%
\end{equation}
where the second term in the brackets corresponds to the renormalization. In
the limit of $M/m\gg1$, as follows from Eq. (\ref{A1}), typical values of
$s_{1}$ are much smaller than typical values of $s_{2}$. We therefore can
neglect $s_{1}$ in the denominator of the first term in Eq. (\ref{E2A}) and
replace $m_{r}$ by $m$. This gives%
\begin{align*}
E^{(2)}  &  =\frac{1}{2}g^{2}\int\frac{d\mathbf{q}}{(2\pi\hbar)^{3}}\left[
\int_{0}^{\infty}ds_{1}A_{M}(s_{1},q)\int_{0}^{\infty}ds_{2}\frac{A_{m}%
(s_{2},q)}{s_{2}}-\frac{2m}{q^{2}}n^{2}\right] \\
&  =\frac{1}{2}g^{2}\int\frac{d\mathbf{q}}{(2\pi\hbar)^{3}}\left[  \int
_{0}^{\infty}ds_{1}A_{M}(s_{1},q)\left(  -\frac{1}{2}\right)  \Pi
_{m}(0,q)-\frac{2m}{q^{2}}n^{2}\right] \\
&  =\frac{1}{2}g^{2}\int\frac{d\mathbf{q}}{(2\pi\hbar)^{3}}\left[  n\left[
\theta(q-2p_{F})+\frac{3}{4}\frac{q}{p_{F}}\left(  1-\frac{q^{2}}{12p_{F}^{2}%
}\right)  \theta(2p_{F}-q)\right]  \left(  -\frac{1}{2}\right)  \Pi
_{m}(0,q)-\frac{2m}{q^{2}}n^{2}\right] \\
&  =\frac{1}{2}g^{2}\int\frac{d\mathbf{q}}{(2\pi\hbar)^{3}}\left\{  n\left[
\theta(q-2p_{F})+\frac{3}{4}\frac{q}{p_{F}}\left(  1-\frac{q^{2}}{12p_{F}^{2}%
}\right)  \theta(2p_{F}-q)\right]  \frac{\nu_{m}}{4}\left[  1+\frac{p_{F}}%
{q}\left(  1-\frac{q^{2}}{4p_{F}^{2}}\right)  \ln\frac{2p_{F}+q}{\left\vert
2p_{F}-q\right\vert }\right]  -\frac{2m}{q^{2}}n^{2}\right\} \\
&  =gn^{2}\frac{9(8\ln2-9)}{140}\frac{ap_{F}}{\pi\hbar}.
\end{align*}
As we see, the final result does not depend on $M$, as it was anticipated above.

\section{Order parameter and single-particle excitations}

We now calculate the order parameter and its temperature dependence. At zero
temperature no quasiparticles are present since both quasiparticle energies
are positive ($E_{\pm}>0$). Then, confining ourselves to second order terms in
$g$, the renormalized gap equation reads:%

\begin{align}
\Delta_{0}(\mathbf{p})  &  =-\frac{2\pi\hbar^{2}a}{m_{r}}\int\frac
{d\mathbf{p}^{\prime}}{(2\pi\hbar)^{3}}\left[  \frac{1}{E_{+}(\mathbf{p}%
^{\prime})+E_{-}(\mathbf{p}^{\prime})}-\frac{2m_{r}}{p^{\prime2}}\right]
\Delta_{0}(\mathbf{p}^{\prime})\nonumber\\
&  -\int\frac{d\mathbf{p}^{\prime}}{(2\pi\hbar)^{3}}\delta V(\mathbf{p}%
,\mathbf{p}^{\prime})\frac{1}{E_{+}(\mathbf{p}^{\prime})+E_{-}(\mathbf{p}%
^{\prime})}\Delta_{0}(\mathbf{p}^{\prime}). \label{eq:rengap_zeroT}%
\end{align}
Strictly speaking, the many-body contribution to the interparticle interaction
$\delta V$ is affected by the superfluid pairing and, therefore, does not
coincide with that of Eq.~(\ref{eq:V2}). However, at zero temperature the
difference is proportional to $\Delta_{0}/\mu_{i}$ and, hence, is
exponentially small. Therefore, we can use Eq.~(\ref{eq:V2}) for $\delta V$ in
Eq.~(\ref{eq:rengap_zeroT}).

The arguments used above for obtaining Eq.~(\ref{TcF}) from
Eq.~(\ref{eq:linear_ren_gap}), can also be applied here, but the large
logarithm \ $\ln(\mu/T_{c})$ should be replaced by $\ln(\mu/\Delta_{0}%
(p_{F}))$.

For the value of the order parameter at the Fermi surface, $\Delta_{0}(p_{F}%
)$, we obtain
\begin{equation}
\Delta_{0}(p_{F})=\left(  \frac{2}{e}\right)  ^{7/3}\exp\left\{  -\frac
{1+\ln4}{3}[f(\kappa)+f(\kappa^{-1})-1]\right\}  \frac{p_{F}^{2}}{4m_{r}}%
\exp\left(  -\frac{1}{\lambda}\right)  . \label{eq:delta0}%
\end{equation}
Comparing Eq.~(\ref{eq:TcGMB}) with Eq.~(\ref{eq:delta0}) we obtain a relation
between $\Delta_{0}(p_{F})$ and $T_{c}$:%

\begin{equation}
T_{c}=\frac{e^{\gamma}}{\pi}\frac{2}{\kappa^{1/2}+\kappa^{-1/2}}\Delta
_{0}(p_{F}). \label{TcDelta}%
\end{equation}
We should emphasize that relation (\ref{eq:delta0}) between the order
parameter and the critical temperature remains valid after taking into account
higher order terms in the gap equation, which is necessary for a large mass
ratio (see the previous section for the discussion of the critical
temperature). The generalization of Eq.~(\ref{eq:rengap_zeroT}) in order to
include the higher order terms repeats the derivation of Eq. (\ref{lingap3})
and the resulting equation reads:
\begin{align}
\Delta(\mathbf{p})  &  =-\frac{2\pi\hbar^{2}a}{m_{r}}\frac{m_{r}^{\ast}}%
{m_{r}}Z_{M}Z_{m}\int\frac{d\mathbf{p}^{\prime}}{(2\pi\hbar)^{3}}\left[
\frac{1}{E_{+}(\mathbf{p}^{\prime})+E_{-}(\mathbf{p}^{\prime})}-\frac{2m_{r}%
}{p^{\prime2}}\right]  \Delta(\mathbf{n}^{\prime}p_{F})\nonumber\\
&  -\int_{p<\Lambda p_{F}}\frac{d\mathbf{p}^{\prime}}{(2\pi\hbar)^{3}}\left[
\delta V(\mathbf{p},\mathbf{n}^{\prime}p_{F})+\delta V^{(3)}(\mathbf{p}%
,\mathbf{n}^{\prime}p_{F})\right]  \frac{1}{E_{+}(\mathbf{p}^{\prime}%
)+E_{-}(\mathbf{p}^{\prime})}\Delta(\mathbf{n}^{\prime}p_{F}).
\label{gapzero3}%
\end{align}
This equation can be solved in a way similar to that of solving
Eq.~(\ref{lingap3}), and the solution is%
\begin{equation}
\Delta_{0}(p_{F})=\pi e^{-\gamma}\frac{p_{F}^{2}}{4m_{r}}\exp\left\{  -\left[
\frac{m_{r}^{\ast}}{m_{r}}Z_{M}Z_{m}\lambda-\nu_{F}\left(  \overline{\delta
V}+\overline{\delta V^{(3)}}\right)  \right]  ^{-1}\right\}  . \label{Delta3}%
\end{equation}
Comparing Eq.~(\ref{Tc3}) with Eq.~(\ref{Delta3}), we immediately obtain
Eq.~(\ref{TcDelta}).

We now analyze the order parameter in the two limiting cases: $M=m$ and $M\gg
m$. In the case of equal masses we have $m_{r}=m/2$ and recover the usual
expression \cite{GMB} for the order parameter from Eq.~(\ref{eq:delta0}) :
\[
\Delta_{0}(p_{F})=\left(  \frac{2}{e}\right)  ^{7/3}\frac{p_{F}^{2}}{2M}%
\exp\left(  -\frac{\pi\hbar}{2p_{F}|a|}\right)  =0.489\,T_{c};\quad M=m.
\]

In the case of $M\gg m$, i.e. $\kappa\gg1$, we have $m_{r}\approx m$ and,
including higher-order contributions, from Eqs.~(\ref{TcDelta})\ and
(\ref{TcLargeratio}) we obtain
\begin{equation}
\Delta_{0}(p_{F})=2^{8/3}e^{-13/6}\frac{p_{F}^{2}}{2\sqrt{Mm}}\exp\left(
-\frac{\pi\hbar}{2p_{F}|a|}-0.011\frac{\left\vert a\right\vert p_{F}}{\hbar
}\frac{M}{m}\right)  =0.882\,\sqrt{\frac{M}{m}}T_{c}\gg T_{c};\quad M\gg m.
\label{Delta0}%
\end{equation}
Note that in this limit the order parameter at the Fermi surface is much
larger than the critical temperature.

In order to analyze the behavior of the order parameter $\Delta$ for
temperatures close to the critical temperature, $(T_{c}-T)\ll T_{c}$, we have
to expand the gap equation (\ref{eq:renorm_gap}) in powers of $\Delta/T_{c}%
\ll1$ and keep the cubic term. The result can be written as%
\[
\Delta(p_{F})\left[  \ln\frac{T_{c}}{T}-\frac{4\kappa}{(1+\kappa)^{2}}%
\frac{7\zeta(3)}{8\pi^{2}}\left(  \frac{\Delta(p_{F})}{T_{c}}\right)
^{2}\right]  =0.
\]
From this equation we obtain%
\begin{equation}
\Delta(p_{F})=\sqrt{\frac{8\pi^{2}}{7\zeta(3)}}\frac{\kappa^{1/2}%
+\kappa^{-1/2}}{2}T_{c}\sqrt{1-\frac{T}{T_{c}}}. \label{DeltaTc}%
\end{equation}
In the limit of equal masses, $\kappa=1$, we reproduce the well-known result
for the temperature dependence of the order parameter. In the opposite limit
of a large mass ratio, $\kappa\gg1$, we find%
\begin{equation}
\Delta(p_{F})=\sqrt{\frac{M}{m}}\sqrt{\frac{8\pi^{2}}{7\zeta(3)}}T_{c}%
\sqrt{1-\frac{T}{T_{c}}}. \label{DeltaTcLargeratio}%
\end{equation}
Note that as well as Eq.~(\ref{Delta0}), the obtained equation (\ref{DeltaTc})
contains a large factor $\sqrt{M/m}$.

\begin{figure}[ptb]
\begin{center}
\includegraphics[width=8cm]{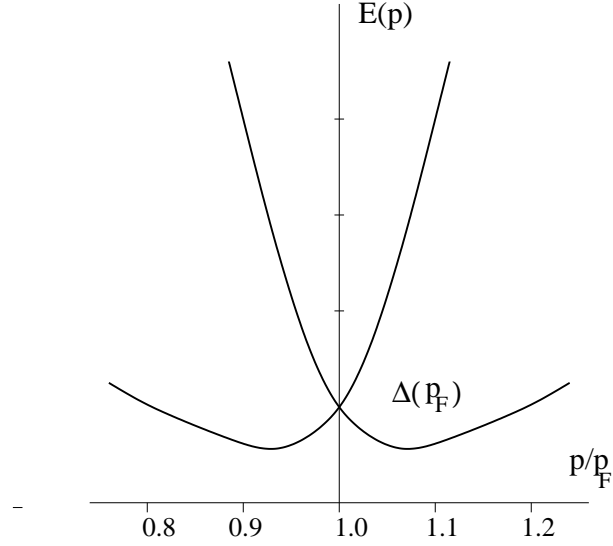}
\end{center}
\caption{Two branches $E_{\pm}(p)$ of single-particle excitations for
$M/m=10$.}%
\label{Fig3}%
\end{figure}

At any temperature the energies of single-particle excitations are given by
(see Eq.~(\ref{eq:dispersion}) and Fig. \ref{Fig3})
\begin{equation}
E_{\pm}(p)=\pm\left(  \frac{p^{2}-p_{F}^{2}}{4m_{-}}\right)  +\sqrt{\left(
\frac{p^{2}-p_{F}^{2}}{4m_{r}}\right)  ^{2}+\Delta^{2}(p_{F})},
\label{eq:dispersion2}%
\end{equation}
where $m_{-}=Mm/(M-m)$. Equation (\ref{eq:dispersion2}) reveals a peculiar feature of
fermionic mixtures with unequal masses of components. For equal masses, the
minimum of $E_{\pm}(\mathbf{p})$ (i.e. the gap) is at the Fermi surface and
equals $\Delta(p_{F})$. The situation for unequal masses is different: the
single-particle excitation energies $E_{\pm}(p)$ reach their minimum values
$E_{\pm\mathrm{min}}=\Delta(p_{F})\sqrt{1-s^{2}}=2\Delta(p_{F})/(\kappa
^{1/2}+\kappa^{-1/2})$ at momenta $p_{\pm}^{\ast2}=p_{F}^{2}\mp2m\kappa
^{1/2}(\kappa-1) \Delta(p_{F})/(\kappa+1)$. For $M/m\gg1$ the corresponding
gap is much smaller than $\Delta(p_{F})$:%
\begin{equation}
E_{\pm\mathrm{min}}(M\gg m)\approx2\Delta(p_{F})\kappa^{-1/2}=2\sqrt{\frac
{m}{M}}\Delta(p_{F})\ll\Delta(p_{F}). \label{Emin}%
\end{equation}
It is interesting to note that the presence of a small factor $\sqrt{m/M}$ in
Eq.~(\ref{Emin}) restores the intuitive picture that the gap in the
single-particle spectrum and the critical temperature are of the same order of
magnitude even in the limit of a large mass ratio. We point out, however, that
in this limit the order parameter on the Fermi surface, being much larger than
the critical temperature, is not equal to the gap in the single-particle
spectrum. This gap is of the order of the critical temperature, and the
low-energy single-particle excitations correspond to momenta different from
the Fermi momentum $p_{F}$. Owing to the former circumstance, one does not
expect any dramatic changes in thermodynamic properties of the system with
increasing the mass ratio $M/m$ to a large value.

\section{Three-body resonances}

Let us now discuss the influence of the three-body physics on the results of
the previous sections. The diagram for the Gorkov-Melikh-Barkhudarov
corrections in Fig. \ref{Fig1} corresponds to collisions between three
particles: two from a Cooper pair and one from the filled Fermi sea. Thus, it
is a three-body process. During this process, however, the three particles
undergo two successive two-body collisions and never appear simultaneously
within the range of the interatomic interaction. The corresponding three-body
wave function vanishes when the hyperspherical radius (see the definition
before Eq.~(\ref{wave_function})) is tending to zero.

In a dilute two-component Fermi gas, real three-body collisions during which
the three colliding particles simultaneously approach each other, are rare. An
additional smallness compared to a Bose gas is provided by the Pauli
principle. Two of the three colliding particles are identical fermions and,
therefore, the wave function of their relative motion should strongly decrease
at small separations. As a result, the contribution of such collisions to the
effective pairing interaction is small and can be neglected. However, for the
case of a large mass ratio $M/m$ the situation is more subtle. If $M/m>13.6$,
two heavy and one light fermions\ can form three-body bound states
\cite{Efimov,Fonseca,PSSJ}. The most interesting case corresponds to the
presence of a weakly bound trimer state because this results in a resonance
$3$-body scattering at low energy. It is not clear that these resonances
should be taken into account when calculating the pairing energy since there
may be nontrivial issues of wave function statistics involved; nevertheless we
shall estimate their possible contribution and leave such issues for further
study \cite{CL}.

\begin{figure}[ptb]
\begin{center}
\includegraphics[width=8cm]{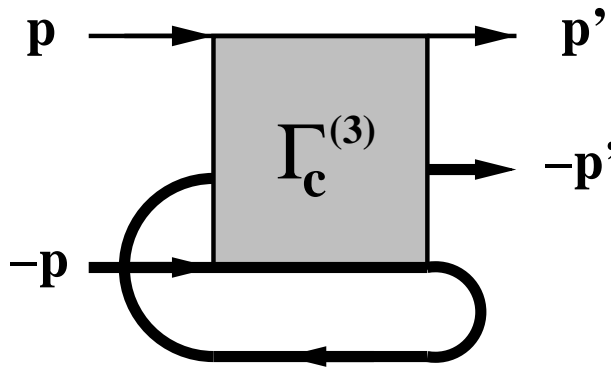}
\end{center}
\caption{The contribution of three-body processes between one light and two
heavy fermions described by the connected three-body vertex $\Gamma_{c}^{(3)}%
$, to the Gorkov-Melikh-Barkhudarov corrections.}%
\label{Fig4}%
\end{figure}

\begin{figure}[ptbptb]
\begin{center}
\includegraphics[width=8cm]{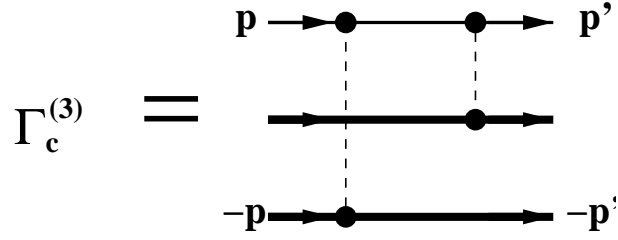}
\end{center}
\caption{The lowest order contribution to the connected three-body vertex.}%
\label{Fig5}%
\end{figure}

To analyze the effect of three-body bound states we note that the contribution
to the Gorkov-Melikh-Barkhudarov corrections in Fig. \ref{Fig1} is part of a
more general contribution involving the connected three-body vertex function
$\Gamma_{c}^{(3)}$(see Fig. \ref{Fig4}). This is a consequence of a general relation between two- and
three-particle Green functions. The quantity $\Gamma_{c}^{(3)}(\left\{
p_{i}\right\}  _{\mathrm{in}},\left\{  p_{i}^{\prime}\right\}  _{\mathrm{out}%
})$ with $p_{i}=(\omega_{i},\mathbf{p}_{i})$, $i=1,2,3$, describes the
scattering of two heavy and one light particle from the initial state with
incoming energy-momenta $p_{i}$ into the final state with outgoing
energy-momenta $p_{i}^{\prime}$. For $\omega_{i}=p_{i}^{2}/2m$ (the mass-shell
condition) the vertex function coincides with the $T$-matrix. By definition,
the connected vertex function $\Gamma_{c}^{(3)}$ does not include three-body
processes in which only two out of the three particles collide (in our case, a
light fermion collides with only one heavy fermion) and, therefore,
$\Gamma_{c}^{(3)}$ is represented only by connected diagrams. The general
three-body vertex function $\Gamma^{(3)}$(see Eq. \ref{Gamma3}) contains all
diagrams including disconnected ones. Those describe processes in which only
two out of the three particles interact with each other. Fig. \ref{Fig5}
shows the simplest contribution to $\Gamma_{c}^{(3)}$ that is
second order and results in the Gorkov-Melikh-Barkhudarov corrections shown in
Fig. \ref{Fig1}.

We consider the case where the size of a three-body bound state is much larger
than $\left\vert a\right\vert $, but much smaller than the average distance
between particles in the gas. Accordingly, the binding energy is much larger
than typical kinetic energies of particles, the Fermi energies $\mu_{i}$. In
this case, the influence of other particles of the gas can be neglected and
the properties of the bound state can be found by solving the three-body
Schr\"{o}dinger equation. Introducing the hyperspherical radius $\rho
=\sqrt{\mathbf{x}^{2}+\mathbf{y}^{2}}$ in the $6$-dimensional space
$(\mathbf{x},\mathbf{y})$, where $\mathbf{x}\sqrt{(2M+m)/4m}=\mathbf{r}%
_{1}-(\mathbf{r}_{2}+\mathbf{r}_{3})/2$ is the distance between the light
fermion and the center of mass of two heavy ones separated from each other by
a distance $\mathbf{y}=\mathbf{r}_{2}-\mathbf{r}_{3}$, the normalized wave
function of a shallow bound state with the binding energy $E_{b}=-\hbar
^{2}/2Mb^{2}$ and the size $b\gg\left\vert a\right\vert $, has the form
\cite{Petrov}:
\begin{equation}
\varphi(\mathbf{\rho})\sim\left(  \frac{m}{M}\right)  ^{3/4}\frac
{1}{\left\vert a\right\vert }\times\left\{
\begin{array}
[c]{l}%
\displaystyle{\frac{\Phi_{1}(\Omega)}{\rho^{2}},\quad\rho\ll\left\vert
a\right\vert }\\
\displaystyle{\frac{\left\vert a\right\vert ^{3}\Phi_{2}(\Omega)}{\rho^{5}%
},\quad a\ll\rho\ll b}%
\end{array}
\right.  . \label{wave_function}%
\end{equation}
Here $\Phi_{1}$ and $\Phi_{2}$ are the functions of hyperangles $\Omega$, and
we do not give explicit expressions for these functions because of their
complexity. For $\rho>b$, the wave function decays exponentially. Note that
the normalization of the wave function (\ref{wave_function}) is determined by
distances $\rho\sim\left\vert a\right\vert $.

Most conveniently the contribution of the bound state to the vertex function
can be found using the three-body Green function $G(\left\{  \mathbf{p}%
_{i}\right\}  ,\left\{  \mathbf{p}_{i}^{\prime}\right\}  ,\omega)$:%
\[
G^{(3)}\left(  \left\{  \mathbf{p}_{i}\right\}  ,\left\{  \mathbf{p}%
_{i}^{\prime}\right\}  ,\omega\right)  =\left\langle \left\{  \mathbf{p}%
_{i}^{\prime}\right\}  \left\vert \frac{1}{\omega-H+i0}\right\vert \left\{
\mathbf{p}_{i}\right\}  \right\rangle ,
\]
where the Hamiltonian $H$ has the form $H=H_{0}+\widehat{V}$ with%
\[
H_{0}=\frac{p_{1}^{2}}{2m}+\frac{1}{2M}\left(  p_{2}^{2}+p_{3}^{2}\right)
\]
and%
\[
\widehat{V}=g\delta(\mathbf{r}_{1}-\mathbf{r}_{2})+g\delta(\mathbf{r}%
_{1}-\mathbf{r}_{3})
\]
in the coordinate representation (index $1$ corresponds to the light fermion
and indices $2$ and $3$ to the heavy ones). The Green function satisfies the
equation%
\[
HG^{(3)}\left(  \left\{  \mathbf{p}_{i}\right\}  ,\left\{  \mathbf{p}%
_{i}^{\prime}\right\}  ,\omega\right)  =\prod\limits_{i=1,2,3}\delta
(\mathbf{p}_{i}-\mathbf{p}_{i}^{\prime}),
\]
which is equivalent to the integral equation%
\begin{equation}
G^{(3)}=G_{0}^{(3)}+G_{0}^{(3)}\widehat{V}G^{(3)}, \label{G3}%
\end{equation}
with $G_{0}^{(3)}=[\omega-p_{1}^{2}/2m-\left(  p_{2}^{2}+p_{3}^{2}\right)
/2M+i0]^{-1}\prod\nolimits_{i}\delta(\mathbf{p}_{i}-\mathbf{p}_{i}^{\prime})$
being the Green function for free particles. This equation can be rewritten in
the form%
\begin{equation}
G^{(3)}=G_{0}^{(3)}+G_{0}^{(3)}\Gamma^{(3)}G_{0}^{(3)}, \label{Gamma3}%
\end{equation}
where we introduce the vertex function $\Gamma^{(3)}$. This function describes
all scattering processes involving three particles, both connected (described
by $\Gamma_{c}^{(3)}$) and disconnected ones (not included in $\Gamma
_{c}^{(3)}$). The vertex function $\Gamma^{(3)}$ obeys the Lipmann-Schwinger
equation%
\begin{equation}
\Gamma^{(3)}=\widehat{V}+\widehat{V}G_{0}^{(3)}\Gamma^{(3)}
\label{Lippmann-Schwinger}%
\end{equation}
and, as it can be seen from Eqs.~(\ref{G3})-(\ref{Lippmann-Schwinger}), is
related to the Green function $G^{(3)}$ as%
\begin{equation}
\Gamma^{(3)}=\widehat{V}+\widehat{V}G^{(3)}\widehat{V}. \label{Gamma3-G3}%
\end{equation}
It is convenient to use the spectral decomposition of the Green function. In
the center-of-mass reference frame, where $\sum_{i}\mathbf{p}_{i}=\sum
_{i}\mathbf{p}_{i}^{\prime}=0$, this decomposition reads:%
\begin{equation}
G^{(3)}\left(  \left\{  \mathbf{p}_{i}\right\}  ,\left\{  \mathbf{p}%
_{i}^{\prime}\right\}  ,\omega\right)  =\sum_{n}\Psi_{n}^{\ast}(\left\{
\mathbf{p}_{i}\right\}  )\frac{1}{\omega-E_{n}+i0}\Psi_{n}(\left\{
\mathbf{p}_{i}^{\prime}\right\}  )+\int d\lambda\Psi_{\lambda}^{(+)\ast
}(\left\{  \mathbf{p}_{i}\right\}  )\frac{1}{\omega-E_{\lambda}+i0}%
\Psi_{\lambda}^{(+)}(\left\{  \mathbf{p}_{i}^{\prime}\right\}  ),
\label{SpectralG3}%
\end{equation}
where the summation is performed over a complete set $\left\{  \Psi_{n}%
,\Psi_{\lambda}^{(+)}\right\}  $ of eigenfunctions of the three-body
Hamiltonian $H$ with eigenenergies $E_{n\text{,}}$ $E_{\lambda}$,
respectively. The eigenfunctions $\Psi_{n}$ correspond to bound states with
energies $E_{n}<0$, and the eigenfunctions $\Psi_{\lambda}^{(+)}$ to
scattering states of three particles with energies $E_{\lambda}>0$. Their
asymptotic behavior at large interparticle distances contains incoming plane
waves with momenta specified by the index $\lambda$, and outgoing (therefore,
index $+$) scattered waves. These eigenfunctions vanish for small
hyperspherical radius $\rho$ and, in particular, they describe the
Gorkov-Melik-Barkhudarov corrections discussed above. On the contrary, the
bound state eigenfunctions are nonzero for small $\rho$ and decay
exponentially for $\rho\rightarrow\infty$.

Eqs.~(\ref{Gamma3-G3}) and (\ref{SpectralG3}) together give the decomposition
of the vertex function $\Gamma^{(3)}$ in terms of the solutions of the
three-body Schr\"{o}dinger equation. Obviously, the bound states contribute
only to the connected vertex function $\Gamma_{c}^{(3)}$. In particular, the
contribution of the bound state $\Psi_{n}$ is:%
\begin{equation}
\delta_{n}\Gamma^{(3)}\left(  \left\{  \mathbf{p}_{i}\right\}  ,\left\{
\mathbf{p}_{i}^{\prime}\right\}  ,\omega\right)  =\delta_{n}\Gamma_{c}%
^{(3)}\left(  \left\{  \mathbf{p}_{i}\right\}  ,\left\{  \mathbf{p}%
_{i}^{\prime}\right\}  ,\omega\right)  =\left[  \widehat{V}\Psi_{n}(\left\{
\mathbf{p}_{i}\right\}  )\right]  ^{\ast}\frac{1}{\omega-E_{n}+i0}\widehat
{V}\Psi_{n}(\left\{  \mathbf{p}_{i}^{\prime}\right\}  ). \label{Gamma3bound}%
\end{equation}
By using the Schr\"{o}dinger equation%
\[
(H_{0}+\widehat{V})\Psi_{n}=E_{n}\Psi_{n},
\]
Eq.~(\ref{Gamma3bound}) can be rewritten in the form%
\begin{equation}
\delta_{n}\Gamma_{c}^{(3)}\left(  \left\{  \mathbf{p}_{i}\right\}  ,\left\{
\mathbf{p}_{i}^{\prime}\right\}  ,\omega\right)  =\Psi_{n}^{\ast}(\left\{
\mathbf{p}_{i}\right\}  )\frac{[E_{n}-E_{0}(\left\{  \mathbf{p}_{i}\right\}
)][E_{n}-E_{0}(\left\{  \mathbf{p}_{i}^{\prime}\right\}  )]}{\omega-E_{n}%
+i0}\Psi_{n}(\left\{  \mathbf{p}_{i}^{\prime}\right\}  ), \label{Gamma3bound1}%
\end{equation}
where $E_{0}(\left\{  \mathbf{p}_{i}\right\}  )=p_{1}^{2}/2m+\left(  p_{2}%
^{2}+p_{3}^{2}\right)  /2M$.

Now we can estimate the contribution of the weakly bound state of one light
and two heavy fermions to the effective interparticle interaction
$V_{\mathrm{eff}}$ between light and heavy fermions with opposite momenta on
the Fermi surface. The analytical expression corresponding to the diagram in
Fig. \ref{Fig4} is%
\begin{equation}
\delta V_{\mathrm{eff}}=\int\frac{d\omega}{2\pi}\int\frac{d\mathbf{q}}%
{(2\pi\hbar)^{3}}\Gamma_{c}^{(3)}\left(  \mathbf{p},\mathbf{q},-\mathbf{p}%
;\mathbf{p}^{\prime},-\mathbf{p}^{\prime},\mathbf{q};\omega\right)  \frac
{1}{\omega-(q^{2}-p_{F}^{2})/2M+i0\mathrm{sign}(q-p_{F})}. \label{effective3}%
\end{equation}
The contribution of the bound state is then obtained by substituting Eq.
(\ref{Gamma3bound1}) into Eq.~(\ref{effective3}) and integrating out the
frequency $\omega$:%
\begin{equation}
\delta_{n}V_{\mathrm{eff}}=\int\frac{d\mathbf{q}}{(2\pi\hbar)^{3}}\theta
(p_{F}-q)\Psi_{n}^{\ast}(\mathbf{p},\mathbf{q},-\mathbf{p})\frac{[E_{n}%
-\mu_{1}-\mu_{2}-q^{2}/2M]^{2}}{(q^{2}-p_{F}^{2})/2M+\mu_{1}+2\mu_{2}%
-E_{n}-q^{2}/[2(2M+m)]}\Psi_{n}(\mathbf{p}^{\prime},-\mathbf{p}^{\prime
},\mathbf{q}), \label{effective3bound}%
\end{equation}
where the last term in the denominator corresponds to the motion of the center
of mass. In the considered case, the binding energy $E_{n}$ is the largest
energy scale ($E_{n}\gg\mu_{1},\mu_{2}$), and $\delta_{n}V_{\mathrm{eff}}$ in
Eq.~(\ref{effective3bound}) can be estimated as%
\begin{equation}
\delta_{n}V_{\mathrm{eff}}\sim(p_{F}/\hbar)^{3}E_{b}\left\vert \Psi
_{n}(0,0,0)\right\vert ^{2}, \label{effective_estimate}%
\end{equation}
where the factor $(p_{F}/\hbar)^{3}$ results from the integration over
$d\mathbf{q}$ and we used the condition $\left\vert a\right\vert p_{F}%
/\hbar\ll1$ to set all momenta in the wave function of the bound state to
zero. After using the wave function from Eq.~(\ref{wave_function}), we
obtain:
\[
\Psi_{n}(0,0,0)\sim(p_{F}/\hbar)\left(  \frac{M}{m}\right)  ^{3/4}(p_{F}%
/\hbar)a^{2}b^{2},
\]
and therefore%
\[
\nu_{F}\delta_{n}V_{\mathrm{eff}}\sim\nu_{F}(p_{F}/\hbar)^{5}E_{b}a^{4}%
b^{4}\left(  \frac{M}{m}\right)  ^{3/2}\sim(p_{F}\left\vert a\right\vert
/\hbar)^{4}(p_{F}b/\hbar)^{2}\sqrt{\frac{M}{m}}.
\]
This result has to be compared with the GM contribution $\nu_{F}%
\overline{\delta V}\sim(p_{F}\left\vert a\right\vert /\hbar)^{2}\ln(M/m)$, and
with the contribution of third-ordrer terms $\nu_{F}\delta V^{(3)}\sim
(p_{F}|a|/\hbar)^{3}M/m$. Under the condition $(p_{F}b/\hbar)<1$ corresponding
to a not too shallow bound state, we find that the contribution of three-body
resonances is small compared to both of them:%
\[
\frac{\nu_{F}\delta_{n}V_{\mathrm{eff}}}{\nu_{F}\overline{\delta V}}\sim
(p_{F}\left\vert a\right\vert /\hbar)^{2}(p_{F}b/\hbar)^{2}\frac{\sqrt{M/m}%
}{\ln(M/m)}\ll1,
\]
and
\[
\frac{\nu_{F}\delta_{n}V_{\mathrm{eff}}}{\nu_{F}\delta V^{(3)}}\sim
(p_{F}\left\vert a\right\vert /\hbar)(p_{F}b/\hbar)^{2}\sqrt{m/M}\ll1.
\]

We thus see that three-body resonances are rather narrow, and their
contribution to the effective interaction can be omitted. So, the results
obtained in the previous sections for the critical temperature, effective
masses, order parameter, and elementary excitations remain unchanged.

\section{Concluding remarks}

We now give an outlook on the physics of attractively interacting mixtures of
heavy and light fermionic atoms in view of the results obtained in this paper.
We have developed a perturbation theory in the BCS limit for the heavy-light
superfluid pairing along the lines proposed by Gorkov and Melik-Barkhudarov
\cite{GMB} and found that for $M/m\gg1$ one has to take into account both the
second-order and third-order contributions. The result for the critical
temperature and order parameter is then quite different from the outcome of
the simple BCS approach. Moreover, the small parameter of the theory is given
by Eq.~(\ref{small_parameter}) and reads: $(p_{F}|a|/\hbar)\ll1$. As we
explained in Section V, this can be seen from the second-order correction to
the fermionic self-energy, which is controlled by the parameter $g^{2}\nu
_{M}\nu_{m}\sim(p_{F}a/\hbar)^{2}M/m$. Therefore, in a mixture of heavy and
light fermions the conventional perturbation theory for the weakly interacting
regime requires much smaller $p_{F}$ (densities) and/or $|a|$ than in the case
of $M\sim m$, where the small parameter is $(p_{F}|a|/\hbar)\ll1$.

\begin{figure}[ptb]
\begin{center}
\includegraphics[width=12cm]{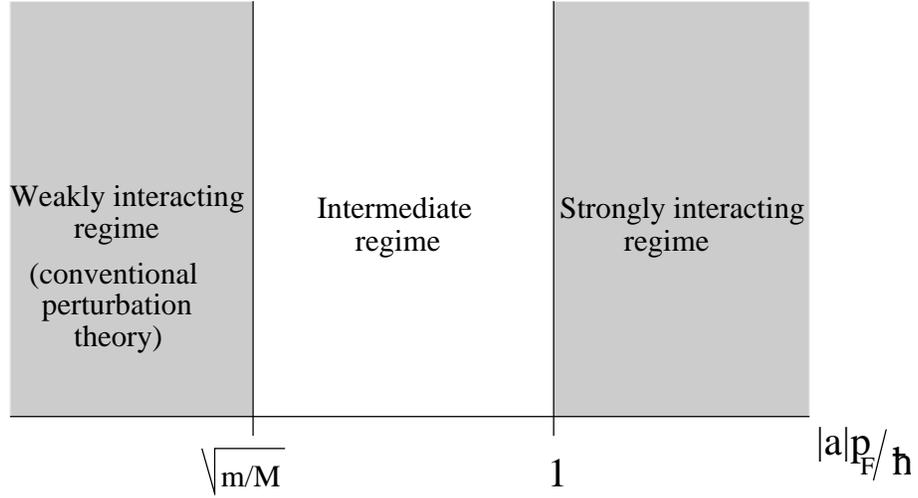}
\end{center}
\caption{Regimes of superfluid pairing for $M/m\gg1$. In the intermediate
regime the conventional perturbation theory is not applicable (see text).}%
\label{C}%
\end{figure}

Let us now discuss the cases of $M=m$ and $M\gg m$ regarding the regimes of
superfluid pairing. For $M=m$ we have the strongly interacting regime for
$(p_{F}|a|/\hbar\gtrsim1$, and the BCS limit for $(p_{F}|a|/\hbar\ll1$. In the
former case the perturbation theory is not applicable and the results are
obtained either by Monte Carlo methods or by adjusting the mean-field theory
to this regime (see \cite{Trento} for review). For $M\gg m$ the situation is
different. As we found, the conventional perturbation theory works well under
the condition $(p_{F}|a|/\hbar)\ll\sqrt{m/M}$ (see Fig.~\ref{C}). For
$(p_{F}|a|/\hbar\gtrsim1$ we have the strongly interacting regime where the
perturbation theory does not work at all. However, we now have a range of
densities and scattering lengths, where $\sqrt{m/M}\ll(p_{F}|a|/\hbar)\ll1$.
In this intermediate regime one can still use Hamiltonian (\ref{H}) and try to
develop a perturbative approach, since the scattering amplitude is much
smaller than the mean interparticle separation. On the other hand, the
conventional perturbation theory does not work for the reasons explained in
Section V. In order to construct a reliable theory one should at least
renormalize the interaction between heavy and light fermions by making an
exact resummation of diagrams containing loops of heavy and light fermions. We
then expect a substantial renormalization of the properties of the superfluid phase.

We thus see that our findings pave a way to revealing novel types of
superfluid pairing in mixtures of attractively interacting ultracold fermionic
atoms with very different masses. An appropriate candidate is a gaseous
mixture of $^{171}$Yb with $^{6}$Li, and one should work out possibilities for
tuning the Li-Yb interaction in this system. Another candidate is a
two-species system of fermionic atoms in an optical lattice with a small
filling factor. The difference in the hopping amplitudes of the species can be
made rather large, which corresponds to a large ratio of the \textquotedblleft
heavy\textquotedblright\ to \textquotedblleft light\textquotedblright%
\ effective mass. For example, in the case of $^{6}$Li-$^{40}$K mixture one
can increase the mass ratio by a factor of $20$ in a lattice with period of
$250$ nm and the tunneling rates $\sim10^{3}$ s$^{-1}$ and $\sim10^{5}$
s$^{-1}$ for K and Li, respectively.

\section*{Acknowledgements}

We acknowledge fruitful discussions with B.L. Altshuler and D.S. Petrov.
Gratefully acknowledged is the hospitality and support of Institute Henri
Poincar\'{e} during the workshop "Quantum Gases" where part of this work has
been done. The work was also supported by the Dutch Foundation FOM, by the
IFRAF Institute, by ANR (grants 05-BLAN-0205 and 06-NANO-014-01), by the
QUDEDIS program of ESF,\ by the Austrian Science Foundation (FWF), and by the
Russian Foundation for Fundamental Research. C.L. acknowledges support from
the EPSRC through the Advanced Fellowship EP/E053033/1. LPTMS is a mixed
research unit No. 8626 of CNRS and Universit\'{e} Paris Sud.


\begin{thebibliography}{99}
\bibitem {foot}According to the Fermi-liquid theory (see, e.g., \cite{LL9}),
quasiparticles in a Fermi system are well-defined only in the vicinity of the
Fermi surface, where their damping is small. Note that quasiparticles in a
weakly interacting gas are well-defined for all momenta, not only for those
close to the Fermi surface. This is because the damping is determined by the
small parameter of the theory and, therefore, is small everywhere. Strictly
speaking, however, the notion of effective masses and $Z$-factors are
attributed to the quasiparticles near the Fermi surface.
\end{thebibliography}
\end{document}